\documentclass[10pt,twocolumn,english,showpacs,amssymb,floatfix,nofootinbib,aps,prd,citesort]{revtex4-1}
\usepackage[T1]{fontenc}
\usepackage[latin9]{inputenc}
\usepackage{color}
\usepackage{babel}
\usepackage{natbib}
\usepackage{units}
\usepackage{amstext}
\usepackage{graphicx}
\usepackage{esint}
\usepackage[unicode=true,
 bookmarks=false,
 breaklinks=false,pdfborder={0 0 1},backref=section,colorlinks=false]
 {hyperref}

\begin{document}

\title{Impact of recent COMPASS data on polarized parton distributions and
structure functions}

\author{M. Salimi-Amiri}
\email{Maryam.salimi.amiri@semnan.ac.ir}

\selectlanguage{english}%

\author{A. Khorramian}
\email{Khorramiana@semnan.ac.ir}

\selectlanguage{english}%

\author{H. Abdolmaleki}
\email{Abdolmaleki@semnan.ac.ir}

\selectlanguage{english}%

\affiliation{Faculty of Physics, Semnan University, 35131-19111, Semnan, Iran}

\author{F. I. Olness}

\affiliation{Department of Physics, Southern Methodist University, Dallas, TX
75275-0175, USA}
\email{olness@smu.edu}

\selectlanguage{english}%

\date{\today}
\begin{abstract}
We perform a new extraction of polarized parton distribution functions
(PPDFs) from the spin structure function experimental data in the
fixed-flavor number scheme (FFNS). In this analysis, we include recent
proton and deuteron spin structure functions obtained by the \texttt{COMPASS}
collaboration. We examine the impact of the \texttt{new COMPASS} proton
and deuteron data on the polarized parton densities and compare with
results from our previous study (KATAO PPDFs), which used the Jacobi
polynomial approach. We find the extracted PPDFs of the proton, neutron,
and deuteron structure functions are in very good agreement with the
experimental data. The results for extracted PPDFs are also compared
with available theoretical models from the literature. 
\end{abstract}

\pacs{13.60.Hb, 12.39.-x, 14.65.Bt
\null\vspace{5.0cm}
}

\maketitle
\tableofcontents{}

\cleardoublepage{}

\section{Introduction\label{Introduction} }


One of the principal goals of Quantum Chromodynamics (QCD) has been
the detailed investigation of the spin structure of the nucleon and
nuclei, as well as the determination of the partonic composition of
their spin projections. The extraction of polarized or spin-dependent
parton distribution functions has been recognized as a longstanding
issue of physical interest \cite{Bass:2004xa,Kuhn:2008sy}, and theoretical
studies on the spin structure of the nucleon have been discussed extensively
in several reviews \cite{Anselmino:1994gn,Lampe:1998eu,Hughes:1999wr,Filippone:2001ux,Altarelli:2009za}.

Determinations of polarized parton distribution functions (PPDFs)
with an estimate of their uncertainties have been presented in multiple
studies \cite{Sidorov:2006fi,Leader:2006xc,Leader:2006ywl,Leader:2005ci,Leader:2005kw,Leader:2010rb,Leader:2014uua,Leader:2004gx,Altarelli:1998nb,Ball:1997sp,Bourrely:1998kg,deFlorian:1997ie,deFlorian:2000bm,deFlorian:2005mw,Gordon:1998xa,Leader:1999rh,Leader:1998nh,Leader:1997kw,Ghosh:2000ys,Gluck:2000dy,Bhalerao:2001rn,Leader:2001kh,Bluemlein:2002be,Goto:1999by,Bourrely:2001du,Forte:2001ph,Altarelli:1996nm,deFlorian:2008mr,Hirai:2008aj,Blumlein:2010rn,deFlorian:2009vb,Leader:2011tm,Leader:2010dx,Leader:2009tr,Ball:2013lla,Nocera:2014uea}.
The variation among these PPDFs sets can be due to a number of factors
including the choice of experimental data sets, the form of the parameterization
and uncertainty calculation, as well as the details of the QCD analysis
such as the treatment of heavy quarks or higher-twist corrections. 

The results from various calculations can lead to a wide range of
expectations for the polarized observables; hence, it is illuminating
to compare the results of different methodologies to the experimental
measurements. In our previous analysis, we performed the detailed
pQCD analysis of PPDFs using the orthogonal Bernstein and Jacobi polynomial
methods at next-to-leading-order (NLO) \cite{Khorramian:2010qa,AtashbarTehrani:2007odq,Khorramian:2004ih}.
Other theoretical studies implementing a QCD analysis on the
spin structure of the nucleon using orthogonal polynomials have been reported in Refs.~\cite{AtashbarTehrani:2013qea,Shahri:2016uzl,Khanpour:2017fey,Khanpour:2017cha}.
Thus, one goal of our investigation is to revisit this topic using
a Mellin moment approach instead of the orthogonal polynomial approach. 

For the present study, we will focus on the polarized structure functions
of the nucleon $g_{1}^{p,n,d}(x,Q^{2})$ which play an important role
in the behavior of polarized parton distribution functions (PPDFs).
Polarized DIS lepton-nucleon scattering has been measured by DESY
\cite{Airapetian:2006vy,Ackerstaff:1997ws,Airapetian:1998wi}, SLAC
\cite{Anthony:1996mw,Abe:1997dp,Abe:1997cx,Abe:1998wq,Anthony:1999rm,Ashman:1989ig,Anthony:2000fn},
\texttt{COMPASS}\cite{Ashman:1987hv,Adeva:1998vv,Alekseev:2010hc,Ageev:2005gh,Alexakhin:2006oza},
CLAS \cite{Dharmawardane:2006zd} and JLAB \cite{Zheng:2004ce}. 

Recently, the \texttt{COMPASS} collaboration \cite{Adolph:2015saz,Adolph:2016myg}
extracted new DIS measurements of the polarized proton and deuteron
structure functions for the region of $0.0035<x<0.575$, $1.03<Q^{2}<96.1$
GeV$^{2}$ and $0.0045<x<0.569$, $1.03<Q^{2}<74.1$ GeV$^{2}$, respectively.
Thus, we will combine the data sets used in Ref.~\cite{Khorramian:2010qa}
with the \texttt{COMPASS16} and \texttt{COMPASS17} data sets to extract
improved polarized structure functions and PPDFs.

The plan of this paper is as follows. In Sec.~II, we review the theoretical
framework and basic formalism of the polarized structure function
analysis based on the inverse Mellin technique. In Sec.~III we outline
the parameterization of PPDFs and the selection of the data sets.
In Sec.~IV, we present the structure functions, PPDFs, and moments
obtained in our fit, and compare these both to our earlier KATAO fit
(using orthogonal polynomials) as well as other results from the literature;
this also includes an evaluation of the impact of the new \texttt{COMPASS}
data sets. Finally, in Sec.~V, we provide a summary and concluding
remarks.

\section{Theoretical Formalism}

The QCD formalism allows us to express the spin dependent nucleon
structure function $g_{1}(x,Q^{2})$ in terms of a Mellin convolution
of the polarized non-singlet $\delta q_{i}^{NS}$, the polarized singlet
$\delta\Sigma$, and the polarized gluon $\delta g$ distributions
with the corresponding Wilson coefficient functions $\delta C_{q}^{NS}$,
$\delta C_{q}^{S}$, and $\delta C_{g}$. The polarized structure function
is then given by \cite{Lampe:1998eu} 
\begin{eqnarray}
g_{1}(x,Q^{2}) & = & \frac{1}{2}\sum_{j=1}^{n_{f}}e_{j}^{2}\left\{ \delta q_{j}^{NS}\otimes\left[1+\frac{\alpha_{s}}{2\pi}\delta C_{q}^{NS}\right]\right.\nonumber \\
 &  & \frac{1}{n_{f}}\delta\Sigma\otimes\left[1+\frac{\alpha_{s}}{2\pi}\delta C_{q}^{S}\right]\nonumber \\
 &  & +\left.\frac{\alpha_{s}}{2\pi}\delta g(x,Q^{2})\otimes\delta C_{g}\right\} \ ,\label{eq:g1dis}
\end{eqnarray}
where $e_{j}$ denotes the charge of the $j$th quark flavor, $n_{f}$
is the number of light flavors, $x$ is the Bjorken variable, $Q^{2}=-q^{2}$
is the four-momentum transfer, and the symbol $\otimes$ denotes the Mellin
convolution. The coefficient functions $\delta C_{i}$ which we use
in the present analysis are calculated in the $\overline{{\rm MS}}$\textendash scheme
at next-to-leading order \cite{Vogelsang:1995vh,Furmanski:1981cw,Bodwin:1989nz,Zijlstra:1993sh};
in particular, we make use of the Pegasus routines \cite{Vogt:2004ns}.
The spin dependent flavor non-singlet distribution $\delta q_{j}^{NS}$
evolves independently, while the spin dependent singlet $\delta\Sigma$
and gluon distributions $\delta g$ are coupled in the QCD evolution.

In the above equation, the polarized non-singlet and singlet PPDFs
are expressed by the individual spin dependent quark flavor contributions
as 
\begin{eqnarray}
\delta\Sigma & = & \sum_{j=1}^{n_{f}}\Big[\delta q_{j}+\delta{\bar{q_{j}}}\Big]~,\\
\delta q_{j}^{NS} & = & \delta q_{j}+\delta{\bar{q_{j}}}-\frac{1}{n_{f}}\delta\Sigma~,
\end{eqnarray}
where $\delta q_{j}$ is the polarized quark distribution function
of the $j$th light flavor.

In our fits, we will take the strong coupling constant $\alpha_{s}(Q_{0}^{2})$
at initial scale $Q_{0}^{2}$ as a free parameter to be fit. The evolution
of the strong coupling constant $\alpha_{s}(Q^{2})$ can be obtained
from the QCD renormalization group equation and is determined by the
$\beta$-function, $\beta(Q^{2})$: 
\begin{eqnarray}
\frac{d\alpha_{s}(Q^{2})}{d\log(Q^{2})}=\beta(Q^{2})=-\beta_{0}\alpha_{s}^{2}(Q^{2})-\beta_{1}\alpha_{s}^{3}(Q^{2})+O(\alpha_{s}^{4})~.\nonumber \\
\label{eq:qcdbeta}
\end{eqnarray}
Here we have expanded the $\beta$-function in powers of $\alpha_{s}$
out to NLO, and the first two coefficients can be computed in the
$\overline{{\rm MS}}$\textendash scheme to be $\beta_{0}=11-\frac{2}{3}n_{f}$
and $\beta_{1}=102-\frac{38}{3}n_{f}$. Thus, given the value of $\alpha_{s}(Q_{0}^{2})$
at the initial scale $Q_{0}^{2}$, we can numerically solve the differential
equation in Eq.(\ref{eq:qcdbeta}) for any $Q^{2}$ scale \cite{Vogt:2004ns}.
For the present analysis, we will work in the FFNS with $n_{f}=3$
light partonic flavors $\{u,d,s\}$.

For our fit, we will use the spin dependent proton, neutron, and deuteron
structure functions. The spin dependent deuteron structure function
${xg}_{1}^{d}(x,Q^{2})$ can be represented in terms of the proton
and neutron structure functions, ${xg}_{1}^{p}(x,Q^{2})$ and ${xg}_{1}^{n}(x,Q^{2})$
using the relation 
\[
{xg}_{1}^{d}(x,Q^{2})=\frac{1}{2}\left(1-\frac{3}{2}\omega_{D}\right)\left[{xg}_{1}^{p}(x,Q^{2})+{xg}_{1}^{n}(x,Q^{2})\right]~,
\]
where $\omega_{D}=0.05\pm0.01$ is the $D$-state wave probability
for the deuteron \cite{Lacombe:1981eg}.

For comparison with the data, we will need to compute the PPDFs and
structure functions at a variety of $Q^{2}$ scales. The evolution
in $Q^{2}$ is performed using the well-known DGLAP collection of
integro-differential evolution equations~\cite{Ahmed:1976ee,Mertig:1995ny}
which can be solved analytically after a conversion from $x$-space
to Mellin $N$-moment space.

The $N$'th Mellin moments of the spin dependent parton densities
$\delta f(x)$ are defined to be 
\begin{eqnarray}
\delta f(N)=\int_{0}^{1}x^{N-1}\delta f(x)~dx~.
\end{eqnarray}
The Mellin transform will decompose the convolution of parton densities
$\delta f(x)$ of Eq.~(\ref{eq:g1dis}) into a product of Mellin
moments: 
\begin{eqnarray*}
[f\otimes g](N)\equiv\int_{0}^{1}dx^{n-1}\int_{x}^{1}\frac{dy}{y}f\left(\frac{x}{y}\right)g(y)=f(N)g(N)~.
\end{eqnarray*}
To invert the Mellin transform, the argument $N$ is analytically
continued to the complex plane. Note that the basic method of solving
the spin dependent non-singlet, singlet, and gluon evolution equations
in Mellin space is reported in the literature in detail \cite{Furmanski:1981cw,Bodwin:1989nz,Gluck:1989ze,Blumlein:1997em}.

The solution of the flavor non-singlet, singlet and gluon evolution
equations at NLO are given by

\begin{eqnarray}
 &  & \delta q_{j}^{{\rm NS}}(N,Q^{2})=\left(\frac{a_{s}}{a_{0}}\right)^{-P_{{\rm NS}}^{(0)}/\beta_{0}}\times\nonumber \\
 &  & \quad\left[1-\frac{1}{\beta_{0}}(a_{s}-a_{0})\left(\delta P_{{\rm NS}}^{(1)}-\frac{\beta_{1}}{\beta_{0}}P_{{\rm NS}}^{(0)}\right)\right]\delta q_{j}^{{\rm NS}}(N,Q_{0}^{2})\;,\nonumber \\
\label{eq:evl1}
\end{eqnarray}
\begin{eqnarray}
\left(\begin{array}{c}
\delta\Sigma(N,Q^{2})\\
\delta g(N,Q^{2})
\end{array}\right) & = & \left[1+a_{s}U_{1}(N)\right]L(N,a_{s},a_{0})\left[1-a_{0}U_{1}(N)\right]\nonumber \\
 &  & \times\left(\begin{array}{c}
\delta\Sigma(N,Q_{0}^{2})\\
\delta g(N,Q_{0}^{2})
\end{array}\right)~,\label{eq:evl2}
\end{eqnarray}
where $a_{s}\equiv a_{s}(Q^{2})$, $a_{0}=a_{s}(Q_{0}^{2})/4\pi$,
$\delta P_{{\rm NS}}^{(0)}$ and $\delta P_{{\rm NS}}^{(1)}$ denote
the LO and NLO non-singlet splitting functions. Here, the matrices $U_{1}$ and $L$ are
evolution matrices, for some details see Ref.~\cite{Blumlein:1997em}.

Given the initial PPDFs at input scale $Q_{0}^{2}$, Eqs. (\ref{eq:evl1})
and (\ref{eq:evl2}) yield the distributions $\delta q^{{\rm NS}}(N,Q^{2})$,
$\delta\Sigma(N,Q^{2})$ and $\delta g(N,Q^{2})$ in Mellin $N$-space
for any scale. We can then transform back to $x$-space to obtain
$\delta f(x)$ by performing a contour integral in the complex plane~\cite{Arbabifar:2013tma}:
\begin{eqnarray}
\delta f(x)=\frac{1}{\pi}\int_{0}^{\infty}dz~{\sf Im}\left[\exp(i\phi)\,x^{-c(z)}\,\delta f[c(z)]\right]\quad,\label{invers}
\end{eqnarray}
where we choose $c(z)=1.1+\rho[\cos(3\pi/4)+i\sin(3\pi/4)]$. The
basic framework of this method is described in the literature~\cite{Arbabifar:2013tma,Bluemlein:2002be,Blumlein:2010rn}.
The resulting $\delta f(x)$ for all PPDFs depends on the initial
value of $\alpha_{s}(Q_{0}^{2})$ and unknown parameters of the spin
dependent parton distributions; we will now discuss our parameterization
form.

\begin{figure}[!tbh]
\centering{}\includegraphics[clip,width=0.49\textwidth]{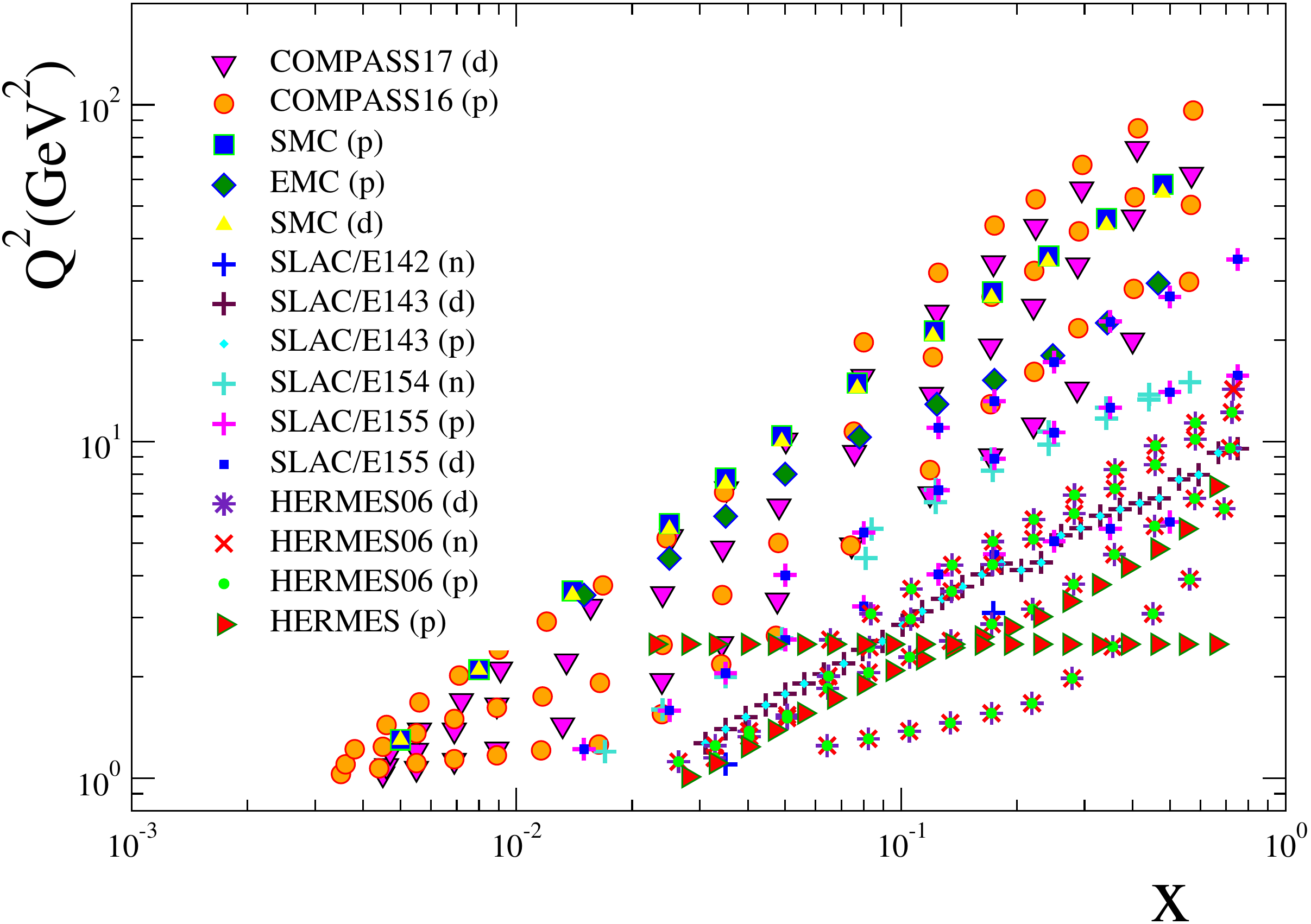}\caption{Experimental data sets used in our fit of proton, deuteron and neutron
structure functions in the $\{x,Q^{2}\}$ plane.{\label{fig:q2x}}}
\end{figure}

\section{Input Parameterization and Data Sets}

\begin{table*}[!t]
\caption{Data sets for polarized DIS structure functions used in our QCD analysis
inclusively covering $0.0035\leqslant x\leqslant0.75$ and $1\leqslant$Q$^{2}\leqslant96.1$
GeV$^{2}$. For each experiment we provide the $x$ and Q$^{2}$ ranges,
the number of data points, and the fitted normalization shifts ${\cal K}_{i}$. }
\label{tab:DISdata} %
\begin{tabular}{ccccccc}
\hline 
{Experiment}  & {Reference}  & \multicolumn{2}{c}{{Data}} & {$x$\textendash Range}  & {Q$^{2}$\textendash Range}  & {${\cal K}_{i}$}\\
\hline 
 &  & Type  & \# data points  &  & (GeV$^{2}$)  & \\
\hline 
HERMES  & \cite{Ackerstaff:1997ws,Airapetian:1998wi}  & DIS (${g_{1}^{p}}$)  & 39  & 0.028-0.66  & 1.01-7.36  & 1.000\\
HERMES06  & \cite{Airapetian:2006vy}  & DIS (${g_{1}^{p}}$)  & 51  & 0.026-0.731  & 1.12-14.29  & 0.999 \\
SLAC/E143  & \cite{Abe:1998wq}  & DIS (${g_{1}^{p}}$)  & 28  & 0.031-0.749  & 1.27-9.52  & 0.999\\
SLAC/E155  & \cite{Anthony:2000fn}  & DIS (${g_{1}^{p}}$)  & 24  & 0.015-0.750  & 1.22-34.72  & 1.023\\
SMC  & \cite{Adeva:1998vv}  & DIS (${g_{1}^{p}}$)  & 12  & 0.005-0.480  & 1.30-58.0  & 1.000\\
EMC  & \cite{Ashman:1987hv}  & DIS (${g_{1}^{p}}$)  & 10  & 0.015-0.466  & 3.50-29.5  & 1.011\\
COMPASS10  & \cite{Alekseev:2010hc}  & DIS (${g_{1}^{p}}$)  & 15  & 0.005-0.568  & 1.10-62.10  & 0.993\\
COMPASS16  & \cite{Adolph:2015saz}  & DIS (${g_{1}^{p}}$)  & 51  & 0.0035-0.575  & 1.03-96.1  & 1.000\\
\textbf{Proton}  &  &  & \textbf{230}  &  &  & \\
\hline 
HERMES06  & \cite{Airapetian:2006vy}  & DIS (${g_{1}^{d}}$)  & 51  & 0.026-0.731  & 1.12-14.29  & 0.997 \\
SLAC/E143  & \cite{Abe:1998wq}  & DIS (${g_{1}^{d}}$)  & 28  & 0.031-0.749  & 1.27-9.52  & 0.998\\
SLAC/E155  & \cite{Anthony:1999rm,Ashman:1989ig}  & DIS (${g_{1}^{d}}$)  & 24  & 0.015-0.750  & 1.22-34.79  & 0.999\\
SMC  & \cite{Adeva:1998vv}  & DIS (${g_{1}^{d}}$)  & 12  & 0.005-0.479  & 1.30-54.80  & 0.999\\
COMPASS17  & \cite{Adolph:2016myg}  & DIS (${g_{1}^{d}}$)  & 43  & 0.0045-0.569  & 1.03-74.1  & 1.001\\
\textbf{Deuteron}  &  &  & \textbf{158}  &  &  & \\
\hline 
HERMES  & \cite{Ackerstaff:1997ws,Airapetian:1998wi}  & DIS (${g_{1}^{n}}$)  & 9  & 0.033-0.464  & 1.22-5.25  & 0.999\\
HERMES06  & \cite{Airapetian:2006vy}  & DIS (${g_{1}^{n}}$)  & 51  & 0.026-0.731  & 1.12-14.29  & 1.000 \\
SLAC/E142  & \cite{Anthony:1996mw}  & DIS (${g_{1}^{n}}$)  & 8  & 0.035-0.466  & 1.10-5.50  & 0.999 \\
SLAC/E154  & \cite{Abe:1997cx}  & DIS (${g_{1}^{n}}$)  & 17  & 0.017-0.564  & 1.20-15.00  & 0.999 \\
\textbf{Neutron}  &  &  & \textbf{85}  &  &  & \\
\hline 
{\textit{Total}}  &  &  & \textbf{473}  &  &  & \\
\hline 
\end{tabular}

\end{table*}

\begin{table*}[!t]
caption{{\small{}Comparison of the parameter values and their
statistical errors at the input scale $Q_{0}^{2}=4$ GeV$^{2}$ in
the different cases: KATAO (Jacobi polynomial method) \cite{Khorramian:2010qa},
Base (without }\texttt{\small{}COMPASS16}{\small{}
and }\texttt{\small{}COMPASS17}{\small{}),
Fit~A (with }\texttt{\small{}COMPASS16}{\small{}),
Fit~B (with }\texttt{\small{}COMPASS16}{\small{}
and }\texttt{\small{}COMPASS17}{\small{})
obtained from the best fit to the data. \label{tab:fit}}}
\begin{tabular}{ccccc}
\hline 
 & \textbf{KATAO (Jacobi Poly.)}  & \textbf{~~~~~~~~~Base~~~~~~~~~}  & \textbf{~~~~~~~~~Fit~A~~~~~~~~~}  & \textbf{~~~~~~~~~Fit~B~~~~~~~~~} \\
\hline 
$\eta_{u_{v}}$  & $~0.928\ (fixed)~$  & $~0.928\ (fixed)~$  & $~0.928\ (fixed)~$  & $~0.928\ (fixed)~$ \\
$\alpha_{u_{v}}$  & $0.535\pm0.022$  & $0.574\pm0.022$  & $0.562\pm0.020$  & $0.570\pm0.019$ \\
$\beta_{u_{v}}$  & $3.222\pm0.085$  & $3.208\pm0.087$  & $3.187\pm0.082$  & $3.207\pm0.079$ \\
$\gamma_{u_{v}}$  & $8.180\ (fixed)$  & $6.527\ (fixed)$  & $6.527\ (fixed)$  & $6.527\ (fixed)$ \\
\hline 
$\eta_{d_{v}}$  & $-0.342\ (fixed)$  & $-0.342\ (fixed)$  & $-0.342\ (fixed)$  & $-0.342\ (fixed)$ \\
$\alpha_{d_{v}}$  & $0.530\pm0.067$  & $0.561\pm0.066$  & $0.591\pm0.063$  & $0.606\pm0.060$ \\
$\beta_{d_{v}}$  & $3.878\pm0.451$  & $3.707\pm0.417$  & $3.895\pm0.415$  & $3.917\pm0.401$ \\
$\gamma_{d_{v}}$  & $4.789\ (fixed)$  & $~3.537\ (fixed)$  & $~3.537\ (fixed)~$  & $~3.537\ (fixed)~$ \\
\hline 
$\eta_{\bar{q}}$  & $-0.054\pm0.029$  & $-0.328\pm0.031$  & $-0.337\pm0.035$  & $-0.309\pm0.018$ \\
$\alpha_{\bar{q}}$  & $0.474\pm0.121$  & $0.500\pm0.125$  & $0.421\pm0.105$  & $0.474\pm0.090$ \\
$\beta_{\bar{q}}$  & $9.310\ (fixed)$  & $10.243\ (fixed)$  & $10.243\ (fixed)$  & $10.243\ (fixed)$ \\
$\gamma_{\bar{q}}$  & $0$  & $0$  & $0$  & $0$ \\
\hline 
$\eta_{g}$  & $0.224\pm0.118$  & $0.231\pm0.102$  & $0.161\pm0.092$  & $0.158\pm0.084$ \\
$\alpha_{g}$  & $2.833\pm0.528$  & $2.737\pm0.456$  & $2.531\pm1.441$  & $2.848\pm0.494$ \\
$\beta_{g}$  & $5.747\ (fixed)$  & $6.323\ (fixed)$  & $6.323\ (fixed)$  & $6.323\ (fixed)$ \\
$\gamma_{g}$  & $0$  & $0$  & $0$  & $0$ \\
\hline 
$\alpha_{s}(Q_{0}^{2})$  & $0.381\pm0.017$  & $0.385\pm0.016$  & $0.388\pm0.015$  & $0.392\pm0.014$ \\
\hline 
$\chi_{{\tt COMPASS16}}^{2}$  & $-$  & $-$  & 32.732  & 33.032 \\
$\chi_{{\tt COMPASS17}}^{2}$  & $-$  & $-$  & $-$  & 28.074\\
$\chi^{2}$  & $273.6$  & $274.8$  & $308.2$  & $337.6$ \\
$d.o.f$  & $370$  & $370$  & $421$  & $464$ \\
$\chi^{2}/d.o.f$  & $0.74$  & $0.74$  & $0.73$  & $0.73$ \\
\hline 
\end{tabular}
\
\end{table*}

\begin{figure*}[!t]
\begin{centering}
\includegraphics[clip,width=0.95\textwidth]{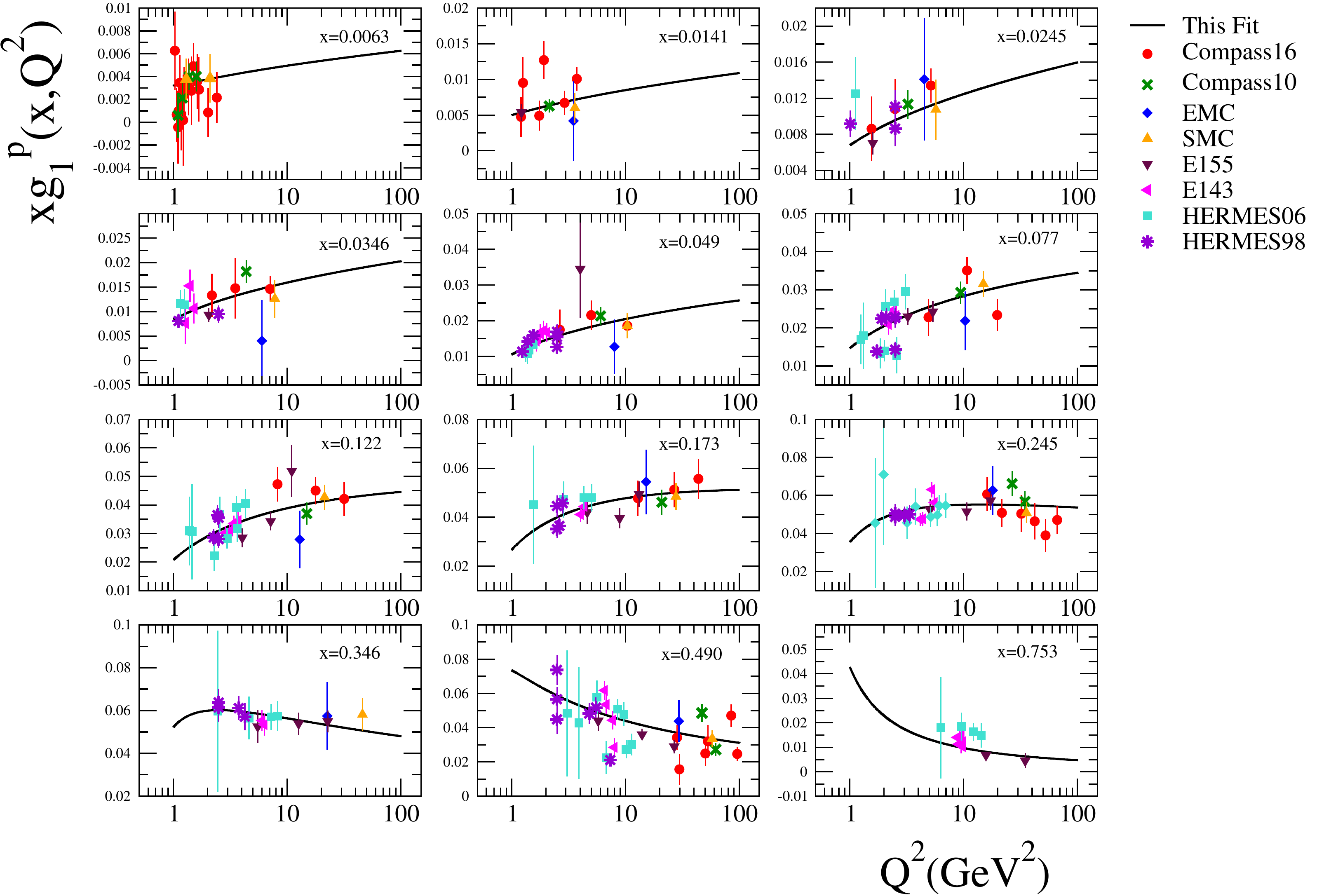} 
\par\end{centering}
\caption{Polarized proton structure function $xg_{1}^{p}(x,Q^{2})$ as a function
of $Q^{2}$ in intervals of $x$ in comparison to the experimental data
of \texttt{COMPASS16} \cite{Adolph:2015saz}, \texttt{COMPASS10} \cite{Alekseev:2010hc},
EMC \cite{Ashman:1987hv}, SMC \cite{Adeva:1998vv}, E155 \cite{Anthony:2000fn},
E143 \cite{Abe:1998wq}, HERMES06 \cite{Airapetian:2006vy}, and HERMES98
\cite{Ackerstaff:1997ws,Airapetian:1998wi}. {\small{}\label{fig:qcdfitp}}}
\end{figure*}

\begin{figure*}[!t]
\begin{centering}
\includegraphics[clip,width=0.95\textwidth]{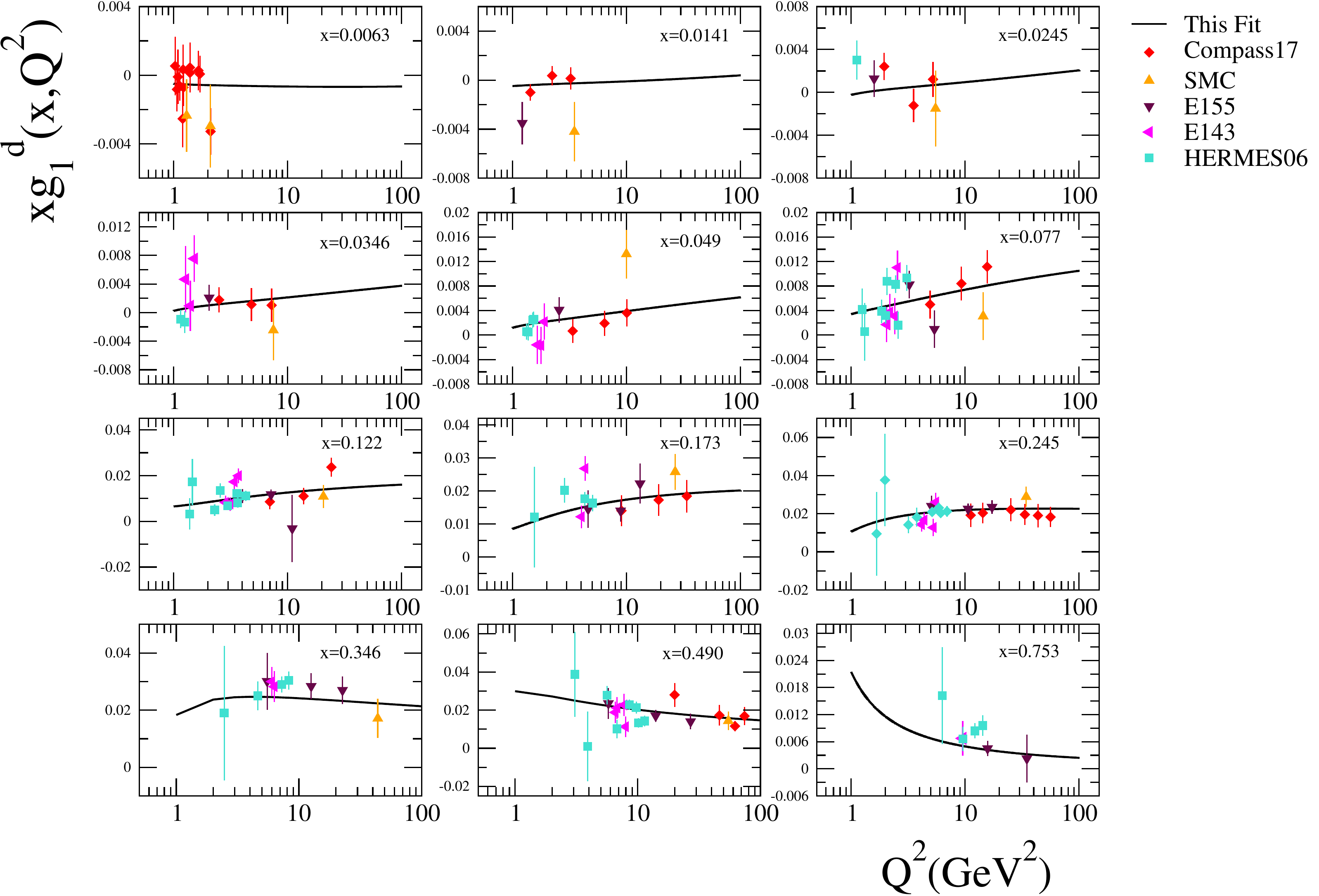} 
\par\end{centering}
\caption{Polarized deuteron structure function $xg_{1}^{d}(x,Q^{2})$ as a function
of $Q^{2}$ in intervals of $x$ in comparison to the experimental data
of \texttt{COMPASS17} \cite{Adolph:2016myg}, SMC \cite{Adeva:1998vv},
E155 \cite{Anthony:1999rm,Ashman:1989ig}, E143 \cite{Abe:1998wq}, and HERMES06 \cite{Airapetian:2006vy}. {\small{}\label{fig:qcdfitd}}}
\end{figure*}

\begin{figure*}[!t]
\begin{centering}
\includegraphics[clip,width=0.95\textwidth]{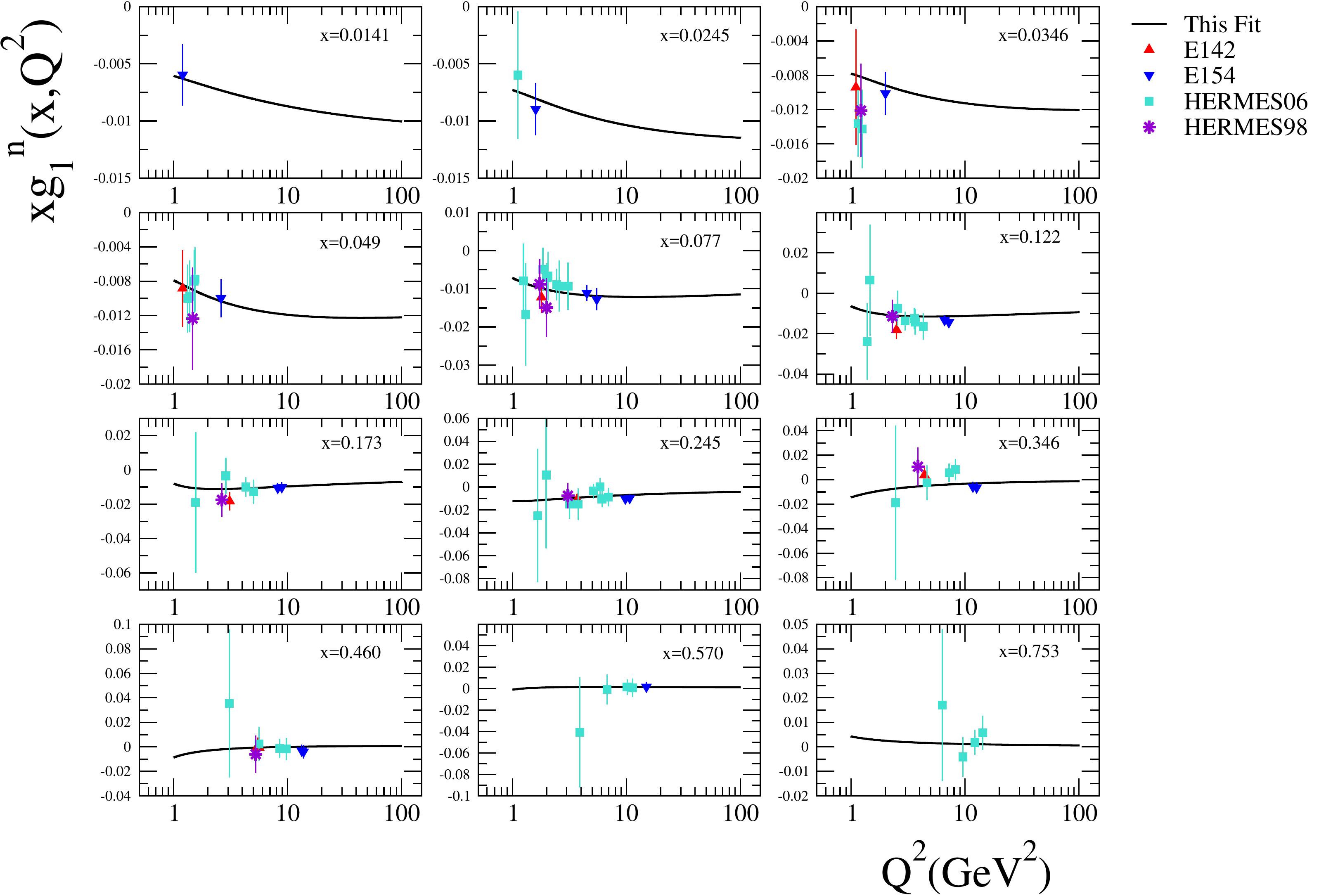} 
\par\end{centering}
\caption{Polarized neutron structure function $xg_{1}^{n}(x,Q^{2})$ as a function
of $Q^{2}$ in intervals of $x$ in comparison to the experimental data
of E142 \cite{Anthony:1996mw}, E154 \cite{Abe:1997cx}, HERMES06
\cite{Airapetian:2006vy}, and HERMES98 \cite{Ackerstaff:1997ws,Airapetian:1998wi}.
{\small{}\label{fig:qcdfitn}}}
\end{figure*}

\begin{figure*}[!htb]
\begin{centering}
\includegraphics[clip,width=0.4\textwidth]{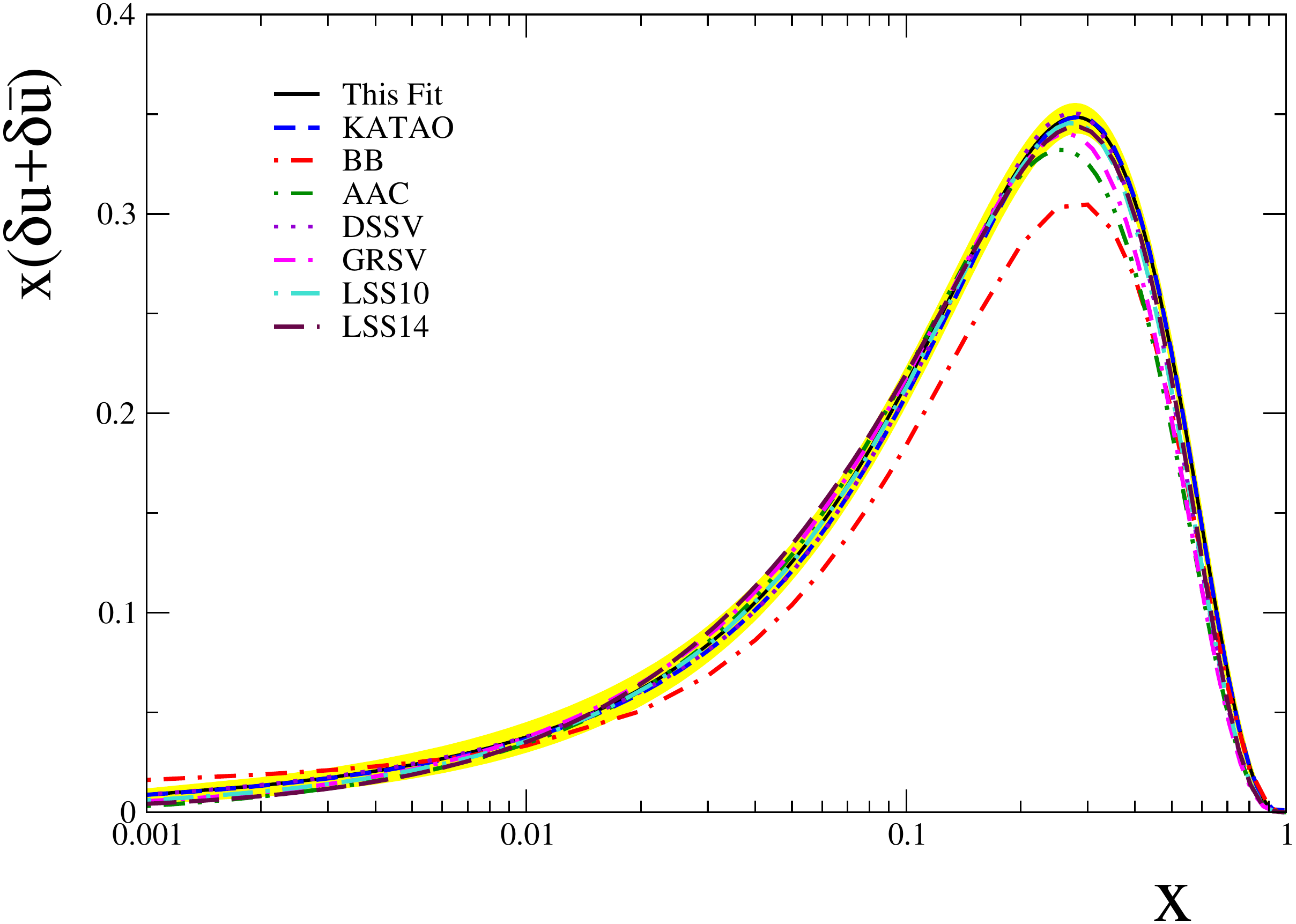} \includegraphics[clip,width=0.4\textwidth]{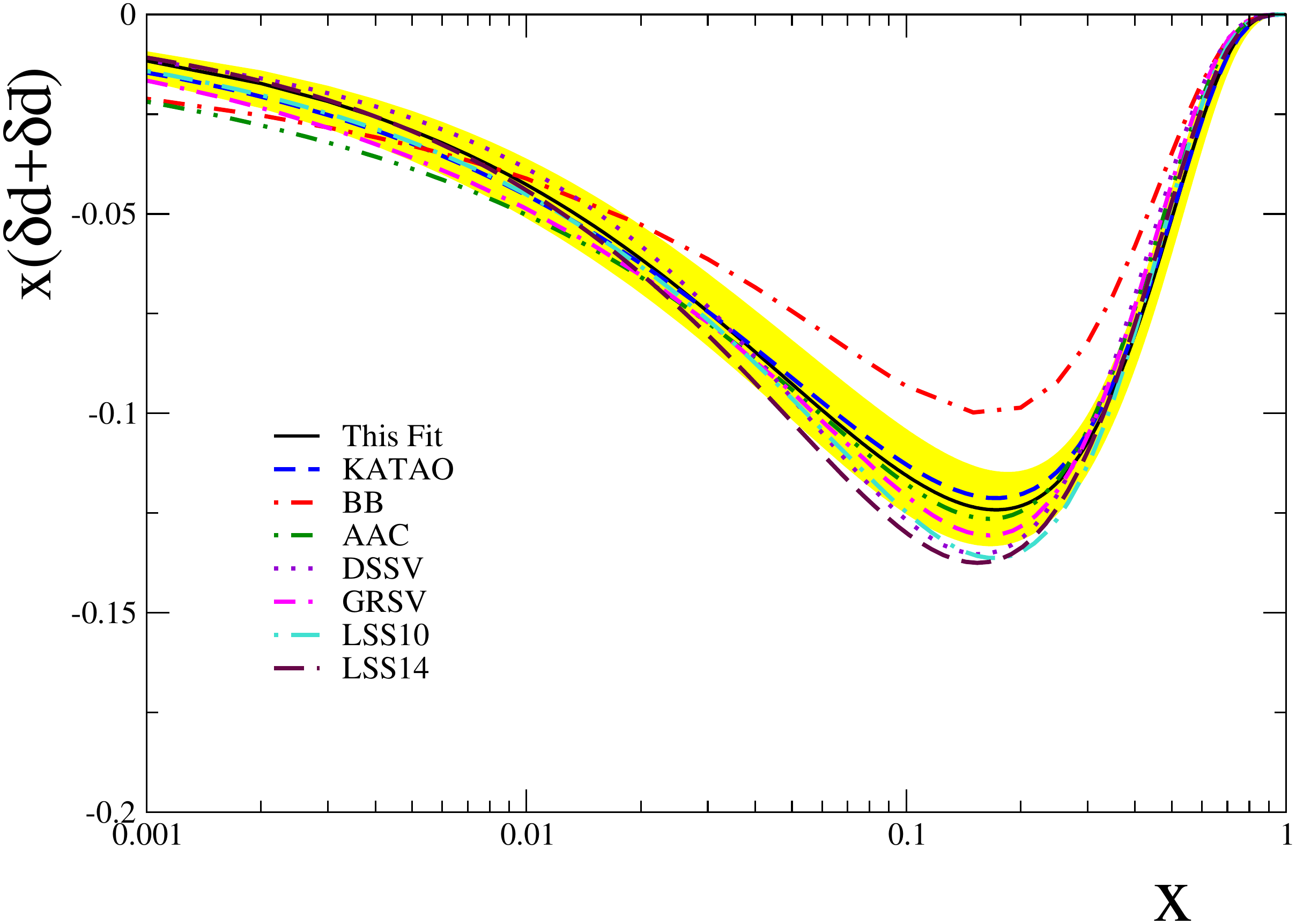} 
\par\end{centering}
\begin{centering}
\includegraphics[clip,width=0.4\textwidth]{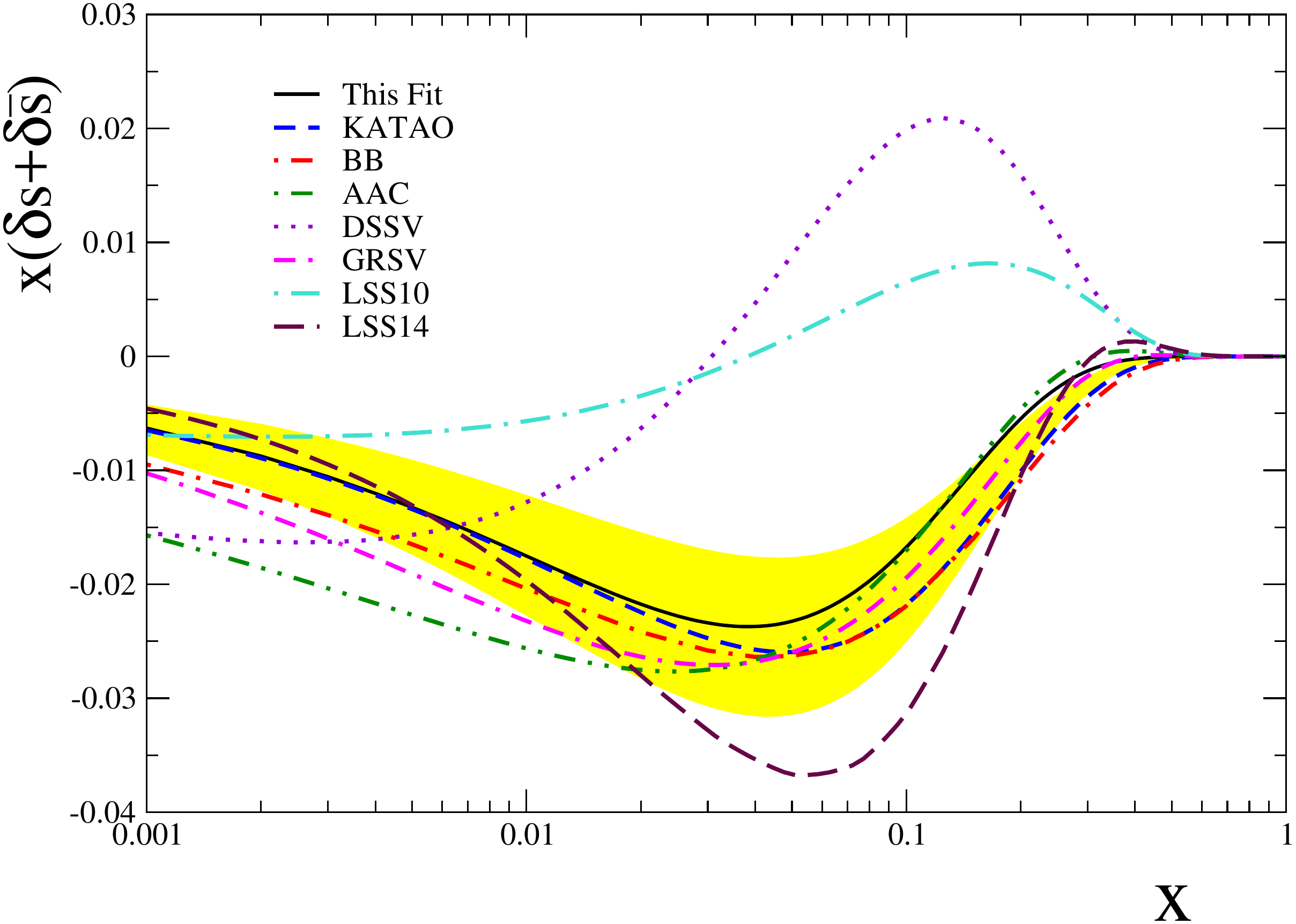} \includegraphics[clip,width=0.4\textwidth]{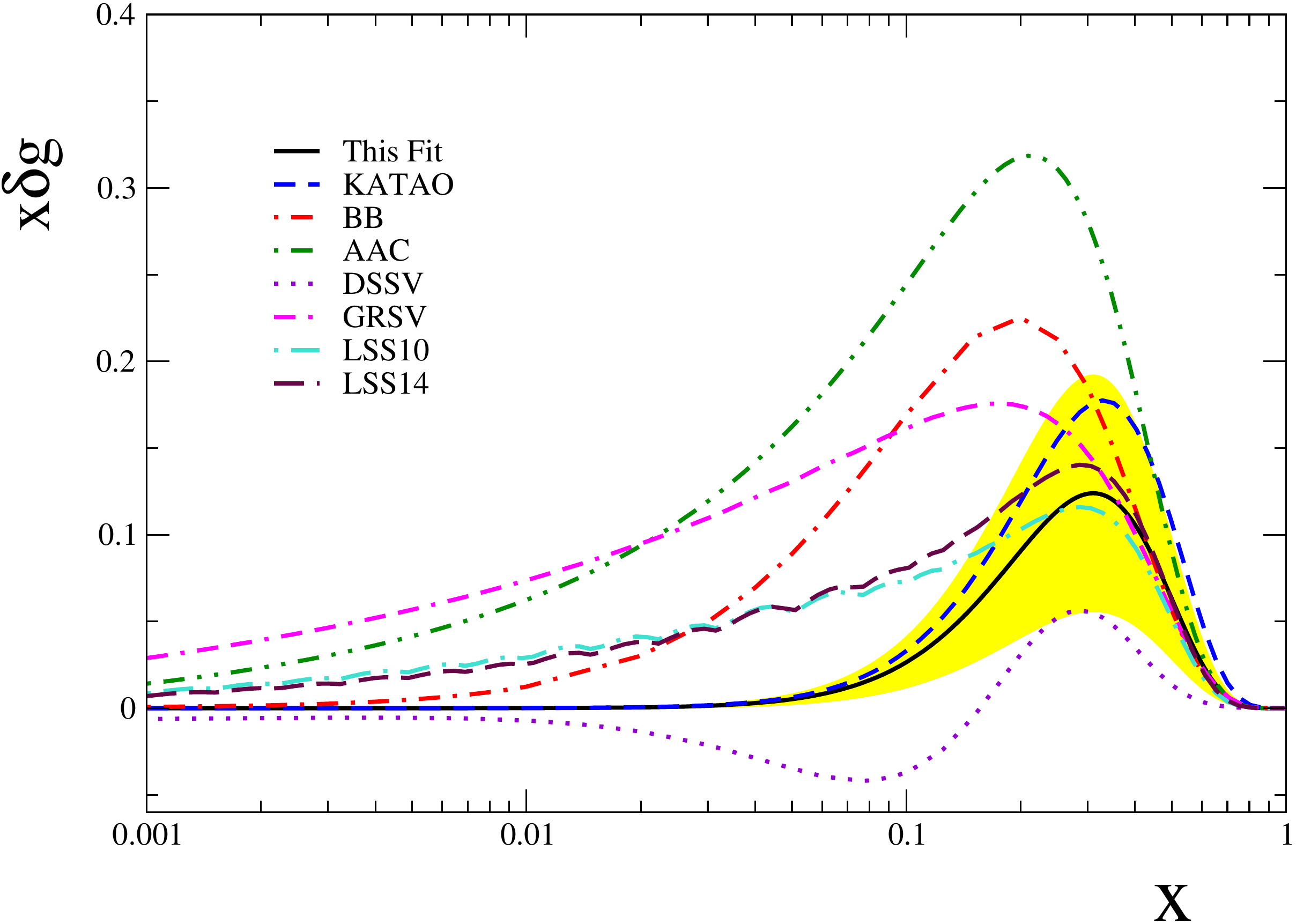} 
\par\end{centering}
\caption{Our results for NLO polarized parton distributions at $Q_{0}{^{2}}$ = 4 GeV$^{2}$. The corresponding PPDFs (the solid lines) are showen with error bands compared to results obtained by KATAO~\cite{Khorramian:2010qa}, BB~\cite{Blumlein:2010rn}, AAC~\cite{Goto:1999by}, DSSV~\cite{deFlorian:2008mr},
GRSV~\cite{Gluck:2000dy} and LSS~\cite{Leader:2010rb,Leader:2014uua}. For clarity, we only present our Fit~B results, labeled here as ``This Fit.''
{\small{}\label{fig:pdf11}}}
\end{figure*}
\begin{figure}[!t]
\begin{centering}
\includegraphics[clip,width=0.23\textwidth]{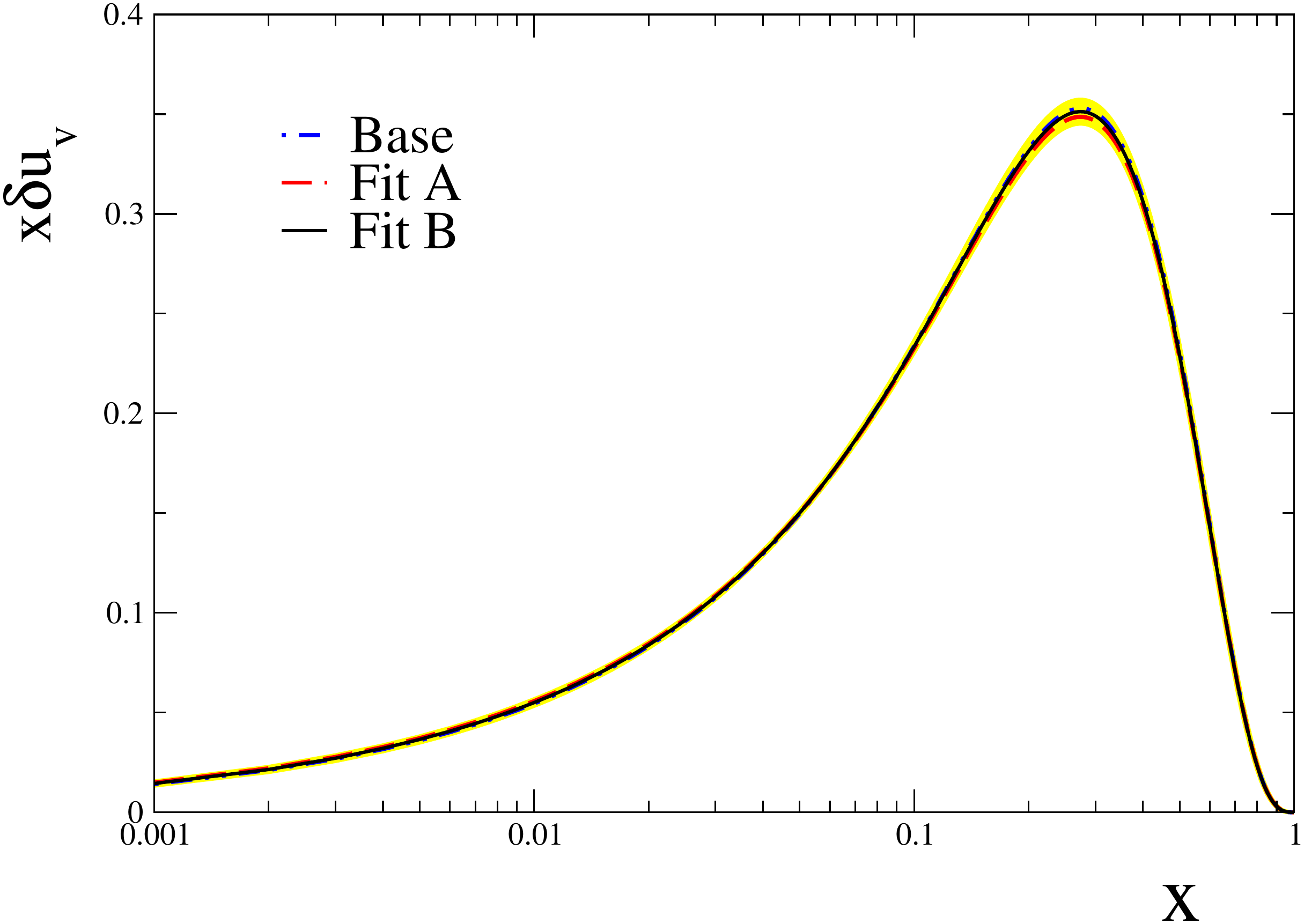}
\hfill{}\includegraphics[clip,width=0.23\textwidth]{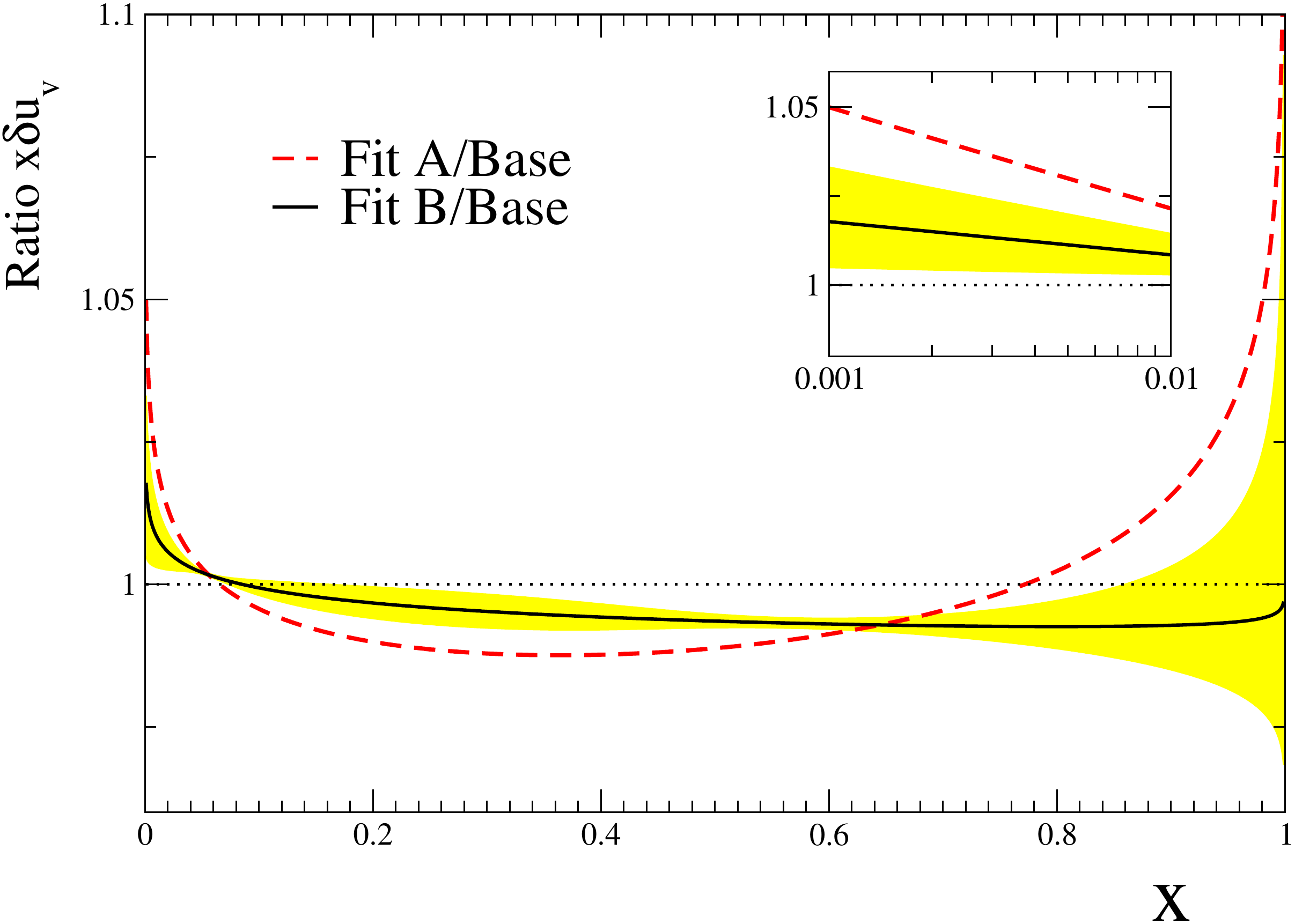} 
\par\end{centering}
\begin{centering}
\includegraphics[clip,width=0.23\textwidth]{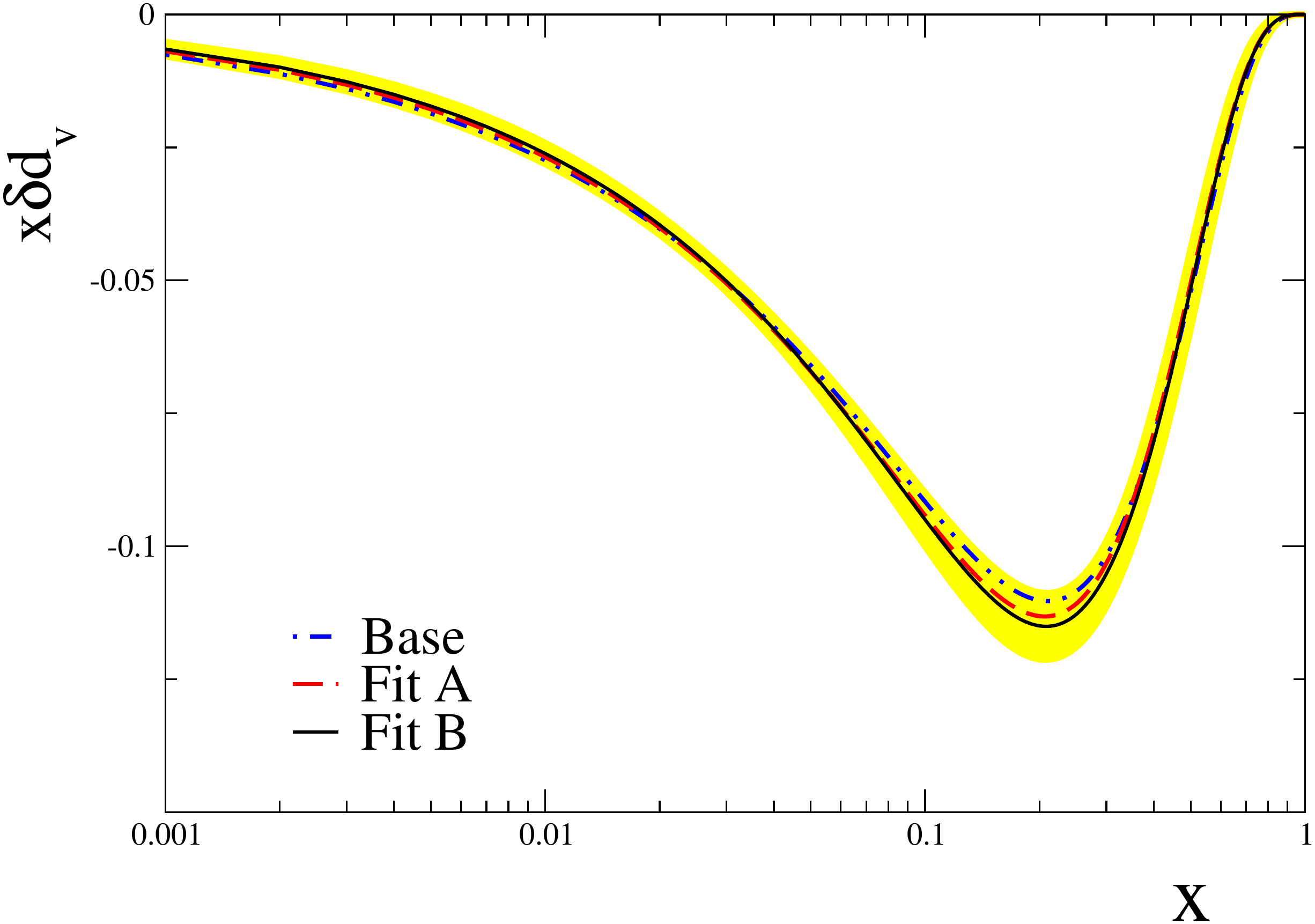}\hfill{}
\includegraphics[clip,width=0.23\textwidth]{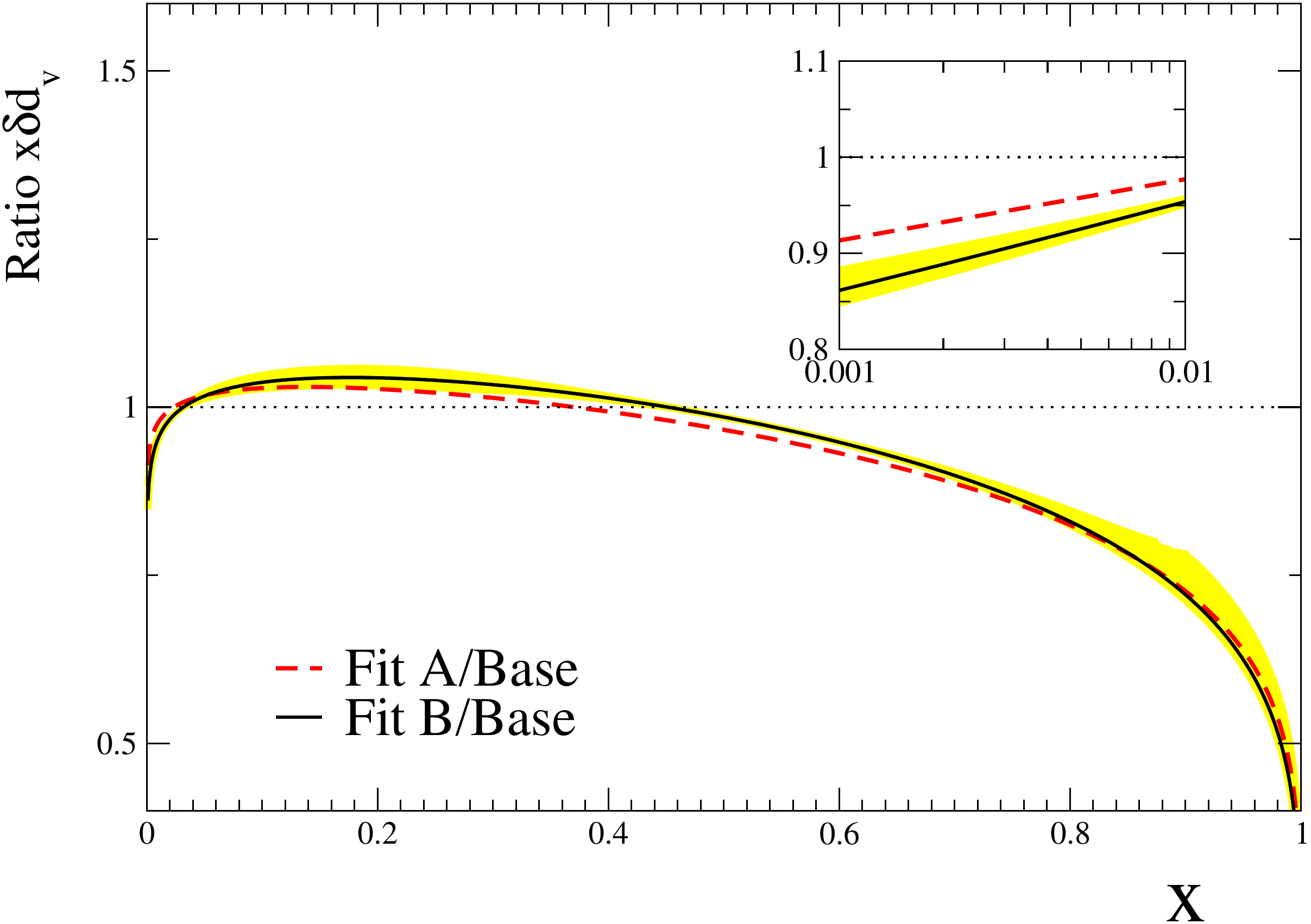} 
\par\end{centering}
\begin{centering}
\includegraphics[clip,width=0.23\textwidth]{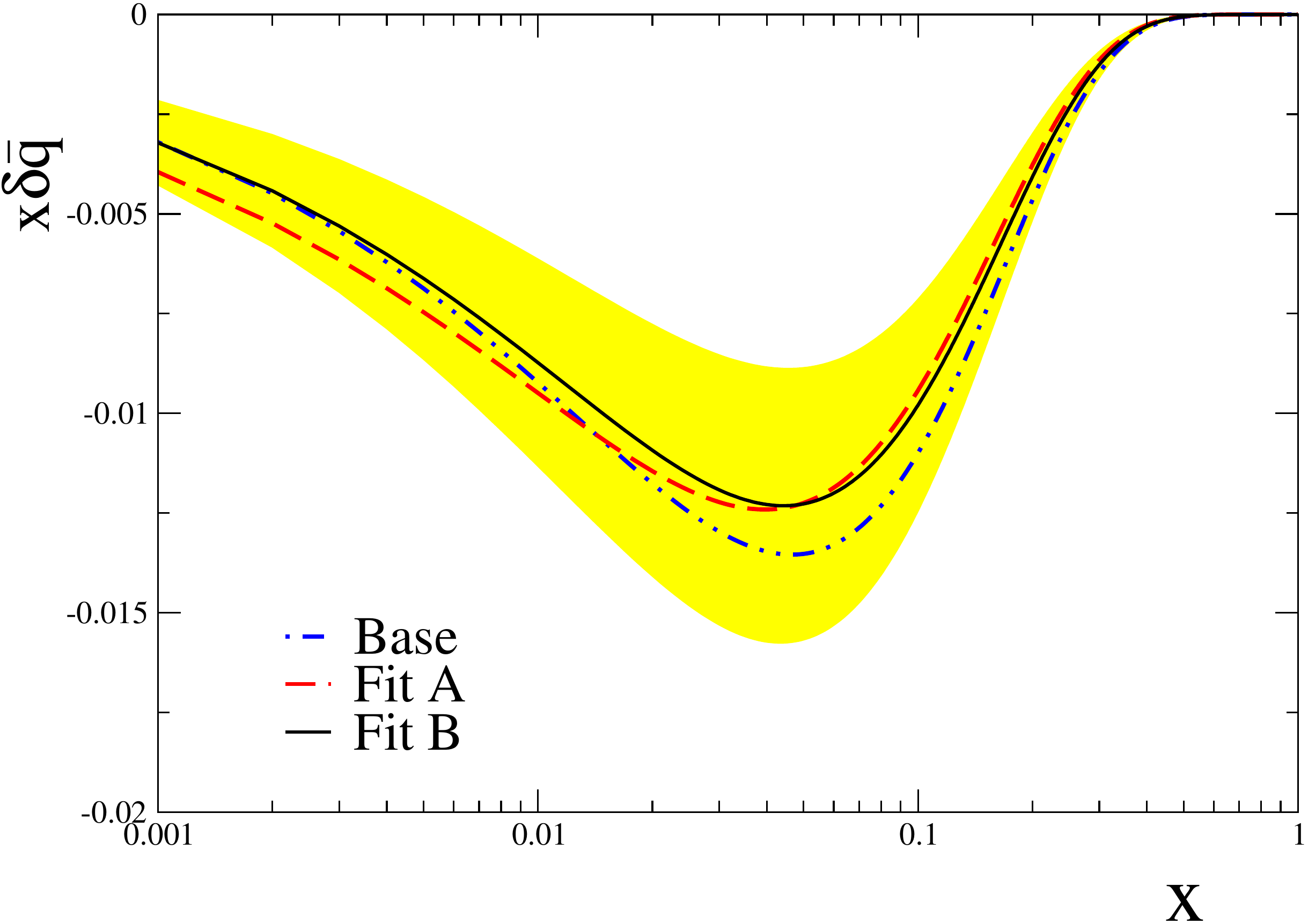}
\hfill{}\includegraphics[clip,width=0.23\textwidth]{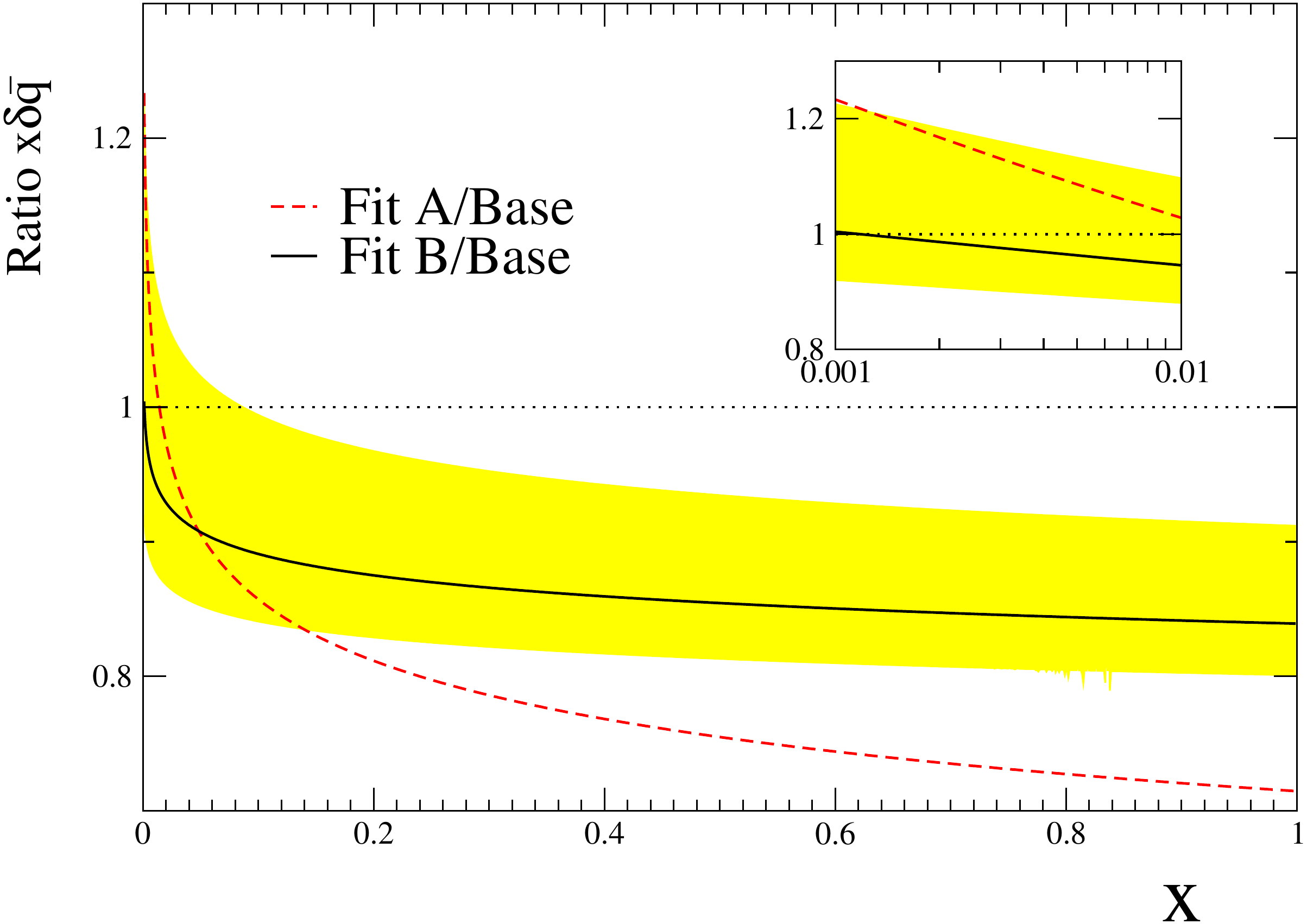} 
\par\end{centering}
\begin{centering}
\includegraphics[clip,width=0.23\textwidth]{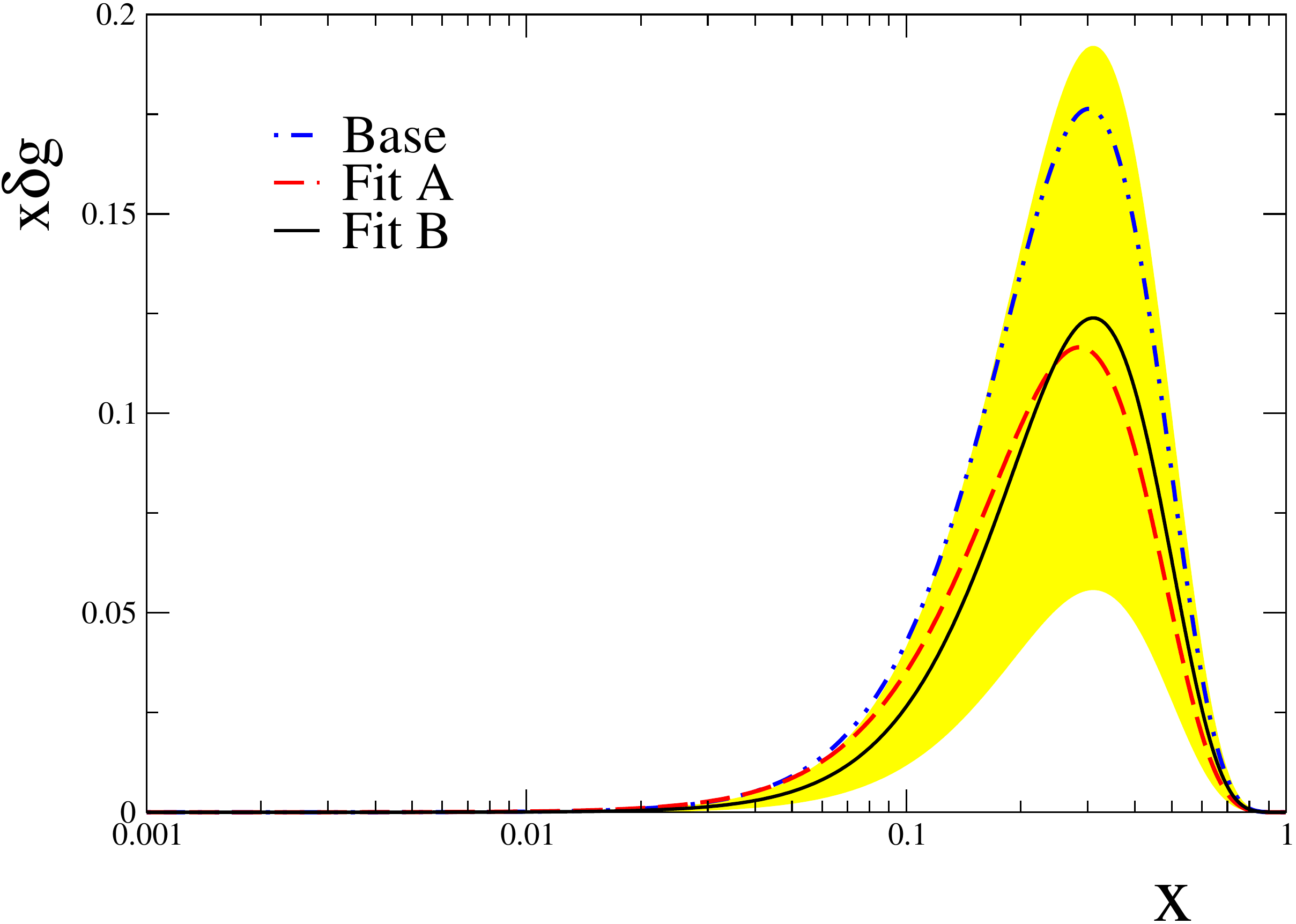}
\hfill{}\includegraphics[clip,width=0.23\textwidth]{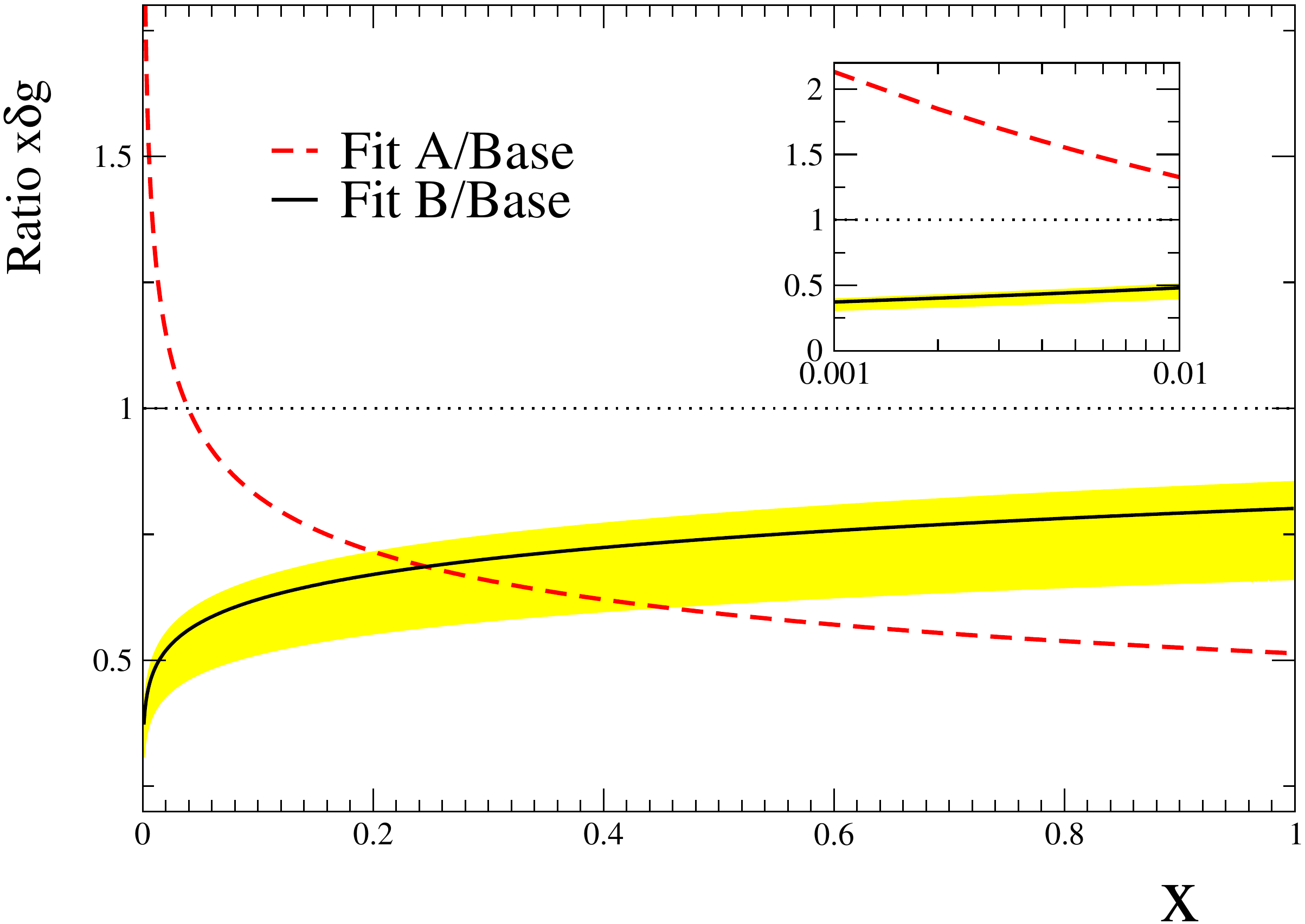} 
\par\end{centering}
\caption{Left panels: polarized parton distributions at $Q_{0}^{2}$ as a function
of $x$ for our different cases: Base (dashed-dotted-dotted), Fit
A (dashed) and Fit~B (solid line) according to Table I. Only the PPDF
error bands for Fit~B (all data) are shown. Right panels: polarized
parton distribution ratios $\delta f$/$\delta f_{Base}$ for Fit
A (dashed) and Fit~B (solid line) to Base obtained from our QCD fits to
the data. The impact of COMPASS data in the low $x$ regions are shown
in the inset plots. {\small{}\label{fig:ratio}}}
\end{figure}


To study the impact of the recent \texttt{COMPASS16} \cite{Adolph:2015saz}
and \texttt{COMPASS17} \cite{Adolph:2016myg} data on the spin dependent
parton distribution functions, we will start by comparing to our previous
KATAO \cite{Khorramian:2010qa} results; hence, our initial parameterization
and $\chi^{2}$ minimization will be based on this work.

\subsection{Parameterization of the polarized parton densities \label{subsec:parm}}

For the parameterization of the spin dependent parton densities in
$x$ space at our initial scale $Q_{0}^{2}=4{\rm \:GeV{}^{2}}$, 
\begin{equation}
x\delta q_{j}(x,Q_{0}^{2})={\eta_{j}{\cal A}_{j}}x^{\alpha_{j}}(1-x)^{\beta_{j}}(1+\gamma_{j}x)\ .\label{eq:parm}
\end{equation}
The free parameters are $\{\eta_{j},\alpha_{j},\beta_{j},\gamma_{j}$\},
and we use the common notation $\delta q_{j}=\{\delta u_{v},\delta d_{v},\delta\bar{q},\delta g\}$
for the partonic flavors up-valence, down-valence, sea, and gluon.
In this functional form, the terms $x^{\alpha_{j}}$ and $(1-x)^{\beta_{j}}$
control the low and large $x$ behavior of the parton densities, respectively.
The $(1+\gamma_{j}x)$ factor controls the intermediate $x$. The
maximal number of parameters which should be fitted for each flavor
component is four $\{\eta_{j},\alpha_{j},\beta_{j},\gamma_{j}$\},
and there are four flavor components $\{\delta u_{v},\delta d_{v},\delta\bar{q},\delta g\}$;
this yields a total of 16 degrees of freedom, but we will introduce
some constraints to reduce the number of free parameters in order
to achieve a stable and reliable minimum.

The ${\cal A}_{j}$ and $\eta_{j}$ parameters are not independent.
Since the first moment of polarized parton densities plays an important
role, the normalization constants ${\cal A}_{j}$ are selected such
that $\eta_{j}$ are the first moments of spin dependent of parton
densities $\delta q_{j}(x,Q_{0}^{2})$; specifically $\eta_{j}=\int_{0}^{1}dx\delta q_{j}(x,Q_{0}^{2})$.
Thus, the normalization factors ${\cal A}_{j}$ can be computed to
be: 
\begin{equation}
\frac{1}{{\cal A}_{j}}=\left(1+\gamma_{j}\frac{\alpha_{j}}{\alpha_{j}+\beta_{j}+1}\right)\,{\cal B}\left(\alpha_{j},\beta_{j}+1\right)\ ,\label{eq:norm}
\end{equation}
where ${\cal B}(m,n)$ is the Euler beta function.

We will presume a $SU(3)$ flavor symmetry such that $\delta\overline{q}\equiv\delta\overline{u}=\delta\overline{d}=\delta s=\delta\overline{s}$.
As we mentioned before, by including only inclusive DIS data in the
QCD fit, it is not possible to separate polarized quarks from polarized
anti-quarks. In fact, inclusive polarized DIS data constrain the total
polarized quarks and anti-quarks combinations.\footnote{In Ref.~\cite{Arbabifar:2013tma}, we reported the results of QCD
analysis using polarized DIS and semi-inclusive DIS (SIDIS) asymmetry
world data, and we extracted the PPDFs considering a light sea-quark
decomposition. }
Thus, we will focus on the PPDF combinations 
  $(\delta{q}+\delta\overline{q})$ as displayed in Fig.~\ref{fig:pdf11}.


Using the above results, we can analytically compute the Mellin-$N$
space transform of the polarized parton densities at the input scale
of $Q_{0}^{2}$: 
\begin{eqnarray}
\delta q_{j}(N,Q_{0}^{2}) & = & \int_{0}^{1}x^{N-1}\:\delta q_{j}(x,Q_{0}^{2})\:dx\nonumber \\
 & = & \eta_{j}{\cal A}_{j}\left(1+\gamma_{j}\:\frac{N-1+\alpha_{j}}{N+\alpha_{j}+\beta_{j}}\right)\nonumber \\
 &  & \times B\left(N-1+\alpha_{j},\beta_{j}+1\right)~.\label{pdfn}
\end{eqnarray}
The first moments of the polarized valence distribution, $\delta u_{v}$
and $\delta d_{v}$, can be fixed by utilizing the parameters $F$
and $D$ as measured in neutron and hyperon $\beta$\textendash decays
\cite{Amsler:2008zzb,Agashe:2014kda}. In fact $q_{3}$ and $q_{8}$ are the non-singlet
combinations of the polarized parton densities:
\begin{eqnarray}
\delta q_{3} & = & (\delta u+\delta\overline{u})-(\delta d+\delta\overline{d})\ ,\\
\delta q_{8} & = & (\delta u+\delta\overline{u})+(\delta d+\delta\overline{d})-2(\delta s+\delta\overline{s})\ .
\end{eqnarray}
The first moments of the above distributions are found to be: 
\begin{eqnarray}
\int_{0}^{1}dx\:\delta q_{3}=\eta_{u_{v}}-\eta_{d_{v}}=F+D\ ,\\
\int_{0}^{1}dx\:\delta q_{8}=\eta_{u_{v}}+\eta_{d_{v}}=3F-D\ .
\end{eqnarray}
Using $F$=0.464$\pm$0.008 and $D$= 0.806$\pm$0.008 from the literature
\cite{Goto:1999by,Aidala:2012mv}, we find the first moments of $\delta u_{v}$
and $\delta d_{v}$ to be $\eta_{u_{v}}=+0.928\pm0.014$ and $\eta_{d_{v}}=-0.342\pm0.018$;
in our QCD fit we will fix $\{\eta_{u_{v}},\eta_{d_{v}}\}$ to these
central values. The first moments of $\delta\overline{q}$ and $\delta g$
do not have prior constraints, and these will be determined in the
fit by the free parameters $\eta_{\bar{q}}$ and $\eta_{g}$.


  The above value for the octet axial charge assumes a good $SU(3)$ symmetry.
  It was noted in Refs.~\cite{Bass:2009ed,Ethier:2017zbq} that this
  symmetry can be broken by about 20\%  which would then yield 
  $F\sim 0.43$ and $D\sim 0.84$, and thus
   $\eta_{u_{v}}\sim +0.865$ and $\eta_{d_{v}}\sim-0.405$. 
  We have also run our fit with these modified values and observed that
  the variation due to these changes is small and well within our PPDF uncertainties.
  %


The factor of $(1+\gamma_{j}x)$ in Eq.~(\ref{eq:parm}) provides
flexibility of the parameterization in the intermediate $x$ region.
This flexibility is beneficial for fitting the the polarized valence
distributions $\delta u_{v},\delta d_{v}$. In contrast, we find that
the parameters $\gamma_{\bar{q}}$ and $\gamma_{g}$ have a very mild
impact on the fit and it is sufficient to set them to zero and remove
these degrees of freedom. (We note that the QCD analysis of polarized SIDIS
data \cite{Arbabifar:2013tma} is sensitive to the $\gamma_{\bar{q}}$
and $\gamma_{g}$ parameters.)

We have now reduced the number of free parameters from 16 to 12. Preliminary
fits indicate that some of the parameters such as $\{\gamma_{u_{v}},\gamma_{d_{v}},\beta_{\bar{q}},\gamma_{g}\}$
are very weakly constrained by the present data set and have very large uncertainties.
In fact, the precision of the data  which we used 
is not high enough to constrain these mentioned parameters sufficiently. We found that, altering them within these uncertainties does not obtain a significant change of $\chi^2$. Therefore we
will also fix the values of these parameters, and we now have a remaining
8 free parameters for the PPDFs in addition to the QCD coupling constant
$\alpha_{s}(Q_{0}^{2})$ to fit from the data.

\subsection{Overview of experimental data set }

The notable advances of the experimental data of inclusive polarized
deep inelastic scattering on nucleons in recent years allows us to
perform an improved QCD analysis of polarized structure functions
in order to discern the spin-dependent partonic structure of the nucleon.
For our analysis, we will include spin structure function data on
protons from HERMES \cite{Ackerstaff:1997ws,Airapetian:1998wi,Airapetian:2006vy},
E143 \cite{Abe:1998wq}, E155 \cite{Anthony:2000fn}, SMC \cite{Adeva:1998vv},
EMC \cite{Ashman:1987hv}, and COMPASS \cite{Alekseev:2010hc,Adolph:2015saz},
on deuterons from HERMES \cite{Airapetian:2006vy}, E143 \cite{Abe:1998wq},
E155 \cite{Anthony:1999rm,Ashman:1989ig}, SMC \cite{Adeva:1998vv}
and COMPASS \cite{Adolph:2016myg},
and on neutrons from HERMES \cite{Ackerstaff:1997ws,Airapetian:1998wi,Airapetian:2006vy},
E142 \cite{Anthony:1996mw} and E154 \cite{Abe:1997cx}. This data
set includes the recent proton data from \texttt{COMPASS16} \cite{Adolph:2015saz}
(51 points), and the recent deuteron data from \texttt{COMPASS17}
\cite{Adolph:2016myg} (43 points). This gives us a total of 473 experimental
data points spanning a kinematic range of 0.0035<$x$<0.75 and 1<$Q^{2}$
<96.1 GeV$^{2}$; these are displayed in Fig.~\ref{fig:q2x}, and
the detailed information and references are summarized in Table~\ref{tab:DISdata}.

In this analysis we will evolve the PPDFs from the initial scale $Q_{0}^{2}=4\ {\rm GeV}{}^{2}$
up to arbitrary scales to compare our theoretical predictions with
the data across the full kinematic range. We construct a global $\chi^{2}$
function using the experimental measurements $g_{1}^{Exp}$, the experimental
uncertainty (statistical and systematic added in quadrature) $\Delta g_{1}^{Exp}$,
and theoretical prediction $g_{1}^{Theory}$. Our $\chi^{2}$ is constructed
as follows: 
\begin{eqnarray}
 & \chi{}_{\mathrm{global}}^{2} & =\sum_{i=1}^{n^{Exp}}w_{i}\chi_{i}^{2}\nonumber \\
 & = & \sum_{i=1}^{n^{Exp}}w_{i}\left[\frac{({\cal K}_{i}-1)^{2}}{(\Delta{\cal K}_{i})^{2}}+\sum_{j=1}^{n^{Data}}\left(\frac{{\cal K}_{i}\:g_{1,j}^{Exp}-g_{1,j}^{Theory}}{{\cal K}_{i}\:\Delta g_{1,j}^{Exp}}\right)^{2}\right],\nonumber \\
\end{eqnarray}
where the $i$-index sums over all experimental data sets, and in
each experimental data set the $j$-index sums over all data points.
We introduce a weight $w_{i}$ which allows us to apply separate weights
to different experimental data sets; for the present analysis we choose
all weights to be unity, $w_{i}=1$.

These data sets include statistical and systematic errors which we
combine in quadrature. There is also a normalization for each experiment
${\cal K}_{i}$ and an associated uncertainty $\Delta{\cal K}_{i}$.
The normalization shifts ${\cal K}_{i}$ are fitted at the start of
our procedure, and then fixed. We present these values in Table~\ref{tab:DISdata},
and find that all the ${\cal K}_{i}$ shifts are less than 1\% except for
a single value; for the SLAC/E155 experiment we find ${\cal K}_{i}=1.024$.

As outlined in Sec.~\ref{subsec:parm}, we have a total of 9 unknown
free parameters: 8 parameters describing the PPDFs at $Q_{0}^{2}$,
and also $\alpha_{s}(Q_{0}^{2})$ as another free parameter. We will
use the CERN library MINUIT package \cite{James:1975dr} to minimize
$\chi^{2}$ by varying the free parameters and obtain a best fit.
We are now ready to extract the polarized parton densities.

\section{Results of the QCD Analysis}

In this section, we will demonstrate how inclusion of the new COMPASS
proton $g_{1}^{p}$ data \cite{Adolph:2015saz} and deuteron $g_{1}^{d}$
data \cite{Adolph:2016myg} influence our PPDFs.

\subsection{Analysis Outline}

\subsubsection{The fits: Base, Fit~A, and Fit~B}

We will divide our analysis into three steps. As a first step, we
perform a fit with all the data of Table~\ref{tab:DISdata} with
the \emph{exception} of the \texttt{COMPASS16} \cite{Adolph:2015saz}
and \texttt{COMPASS17} \cite{Adolph:2016myg} experimental data; this
totals 379 data points, and we identify this as our ``Base'' fit.
We then include the \texttt{COMPASS16} proton data, and this is our
``Fit~A'' which contains 430 data. Finally, we include the \texttt{COMPASS17}
deuteron data, and this is our ``Fit~B'' with the full 473 data
points. As Fit~B contains the complete data set, we will use this
for comparisons in Figs.~\ref{fig:qcdfitp},  \ref{fig:qcdfitd}, \ref{fig:qcdfitn}, and \ref{fig:pdf11} where it is identified as
``This Fit.''

In Table~\ref{tab:fit}, the final values of the fit parameters for
the different data sets are summarized. We find that $\chi^{2}/d.o.f$ is
less than unity in all cases indicating a good quality of fit. Additionally,
our fits compare well with our previous KATAO analysis where we find
$\chi^{2}/d.o.f$=273.6/370.

\subsection{Structure Functions and PPDFs}

\subsubsection{The $xg_{1}^{N}$ Structure Functions vs. $Q^{2}$}

We will begin with the comparison of the $xg_{1}^{N}$ structure functions
as this is the primary input to our fit. In Figs.~\ref{fig:qcdfitp},
\ref{fig:qcdfitd}, and \ref{fig:qcdfitn}, we display the comparison
of our theoretical predictions with the structure function data for
$xg_{1}^{p}$, $xg_{1}^{d}$ and $xg_{1}^{n}$, respectively. The
figures are given as a function of $Q^{2}$ at different values of
$x$ and are compared to all of the experimental data that we used
in the present analysis. The theoretical predictions are in good agreement
with the experimental measurements across the fill $x$-range. In
the following sections, we will investigate the impact of the new COMPASS
measurements on the central values of the PPDFs and their uncertainties.

\subsubsection{The Polarized PDFs (PPDFs)}

Next we turn to the PPDFs themselves. Figure \ref{fig:pdf11} displays
the extracted $x(\delta u+\delta\bar{u})(x)$, $x(\delta d+\delta\bar{d})(x)$, $x(\delta s+\delta\bar{s})(x)$, and $x\delta g(x)$ PPDFs with their associated uncertainties
as compared with various other determinations from the literature \cite{Goto:1999by,Gluck:2000dy,Leader:2010rb,Blumlein:2010rn,deFlorian:2008mr,Khorramian:2010qa}.

We derive the uncertainties of the polarized parton distributions
for the different polarized observables using the covariance matrix
elements of the QCD fit.

Examining Fig.~\ref{fig:pdf11} we find that the spread of results for
the $x(\delta u+\delta \bar u)$ distribution is comparatively narrow indicating
this flavor component is well constrained. The results of ``Fit~B''
are comparable to our previous analysis using the Jacobi polynomial
expansion method (KATAO) \cite{Khorramian:2010qa}, as well as many
of the other results from the literature. Our results are slightly
larger than those of BB in the larger $x$ region ($x\sim0.2$).
The $x(\delta d+\delta \bar d)$ distribution is also comparatively narrow suggesting
this too is well constrained. Again, our results of ``Fit~B'' are
generally comparable to the other results from the literature, with
``Fit~B'' yielding a slightly larger $x(\delta d+\delta \bar d)$ than BB in the region $x\sim0.1$.
For the $x(\delta s+\delta \bar s)$ distributions (or $2 \delta \bar q$ in our notation), we find a broader spread
of both our results (``Fit~B'' and KATAO) and the other fits from
the literature suggesting this component is less constrained. Specifically,
``Fit~B'' roughly coincides with many of the other predictions,
but the DSSV and LSS10 results yield a changes sign as a function of $x$  and LSS14 yields a larger result.
Of all the components we examine, clearly the gluon distribution $x\delta g$
has the widest spread of predictions and the greatest uncertainty.
``Fit~B'' is similar to the KATAO results, but yields a smaller
result in the region $x\sim0.3$; compared to the other curves, these
results generally give a smaller $x\delta g$ than the other predictions.
In particular, in the region $x\sim0.1$ AAC give the largest result
and DSSV gives a negative results. Clearly, the $x\delta g$ distribution
leaves much room for improvement and it will be interesting to see
which predictions are favored by future data sets. Presumably, the
choice of data sets (such as SIDIS) may contribute to these differences.

\subsubsection{Comparison of \{Base,Fit~A, Fit~B\} on PPDFs}

Since it is the new COMPASS data on $xg_{1}^{p}$ and $xg_{1}^{d}$
that represent the important new additions to our data set, we want
to focus on the variations among our fits: \{Base, Fit~A, Fit~B\}.

To investigate the specific impact of \texttt{COMPASS16} and \texttt{COMPASS17}
data sets, we compare our results for our individual fits: ``Base''
(without including \texttt{COMPASS16} and \texttt{COMPASS17}), ``Fit
A'' (including \texttt{COMPASS16}) and ``Fit~B'' (including \texttt{COMPASS16}
and \texttt{COMPASS17}). These results are shown in Fig.~\ref{fig:ratio}
where we have displayed both the absolute value of the PPDFs and also
the ratio compared to our base fit.

As suggested by the results of Fig.~\ref{fig:pdf11}, in Fig.~\ref{fig:ratio} we
find that $x\delta u_{v}(x)$ and $x\delta d_{v}(x)$ appear to be
strongly constrained with little variation among the separate fits.
Specifically, the variation is on the order of a percent except for
the region at large $x$ where the PPDFs vanish and there are no data
constraints. 

In contrast, $x\delta\bar{q}(x)$ and $x\delta g(x)$ do display some
differences amount the fits due to the addition of the COMPASS data;
the variations of ``Fit~A'' and ``Fit~B'' of Fig.~\ref{fig:ratio}
are quite similar, and these differ from the ``Base'' fit. The $x\delta\bar{q}(x)$
function displays some variation in the small $x$ region $\lesssim10^{-1}$
while the variation of $x\delta g(x)$ function is generally at larger
$x\gtrsim10^{-1}$; again, the very large $x$ region should be discounted
as before. 

\subsubsection{COMPASS $xg_{1}^{N}$ Structure Functions vs. $x$}

\begin{figure}[!tbh]
\includegraphics[clip,width=0.45\textwidth]{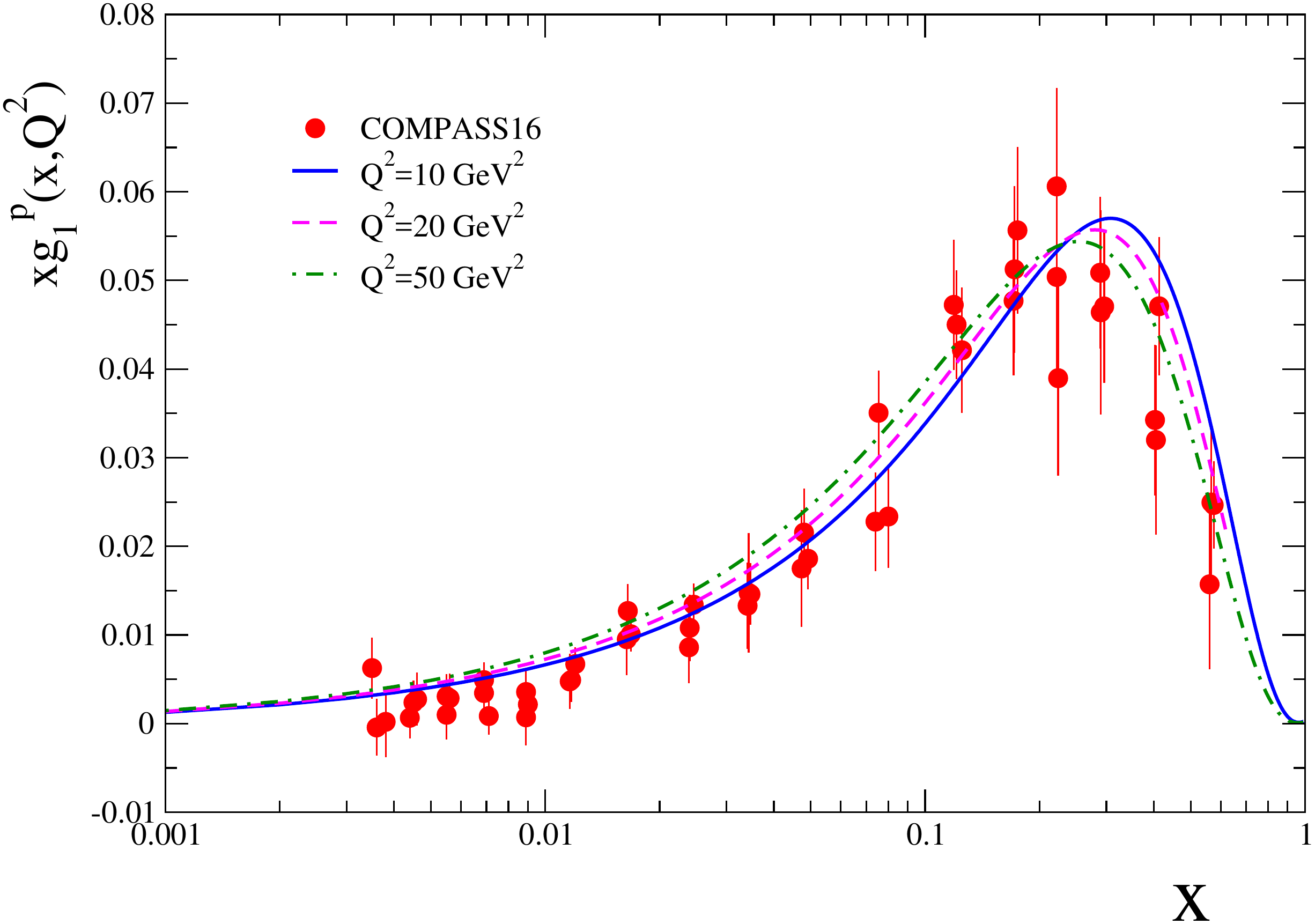}\caption{The \texttt{COMPASS16} \cite{Adolph:2015saz} data for the proton
structure function $xg_{1}^{p}(x,Q^{2})$ compared with our NLO results
calculated at $Q^{2}$ = 10, 20, 50 GeV$^{2}$.{\small{}\label{fig:xg1p}} }
\end{figure}

\begin{figure}[!tbh]
\includegraphics[clip,width=0.45\textwidth]{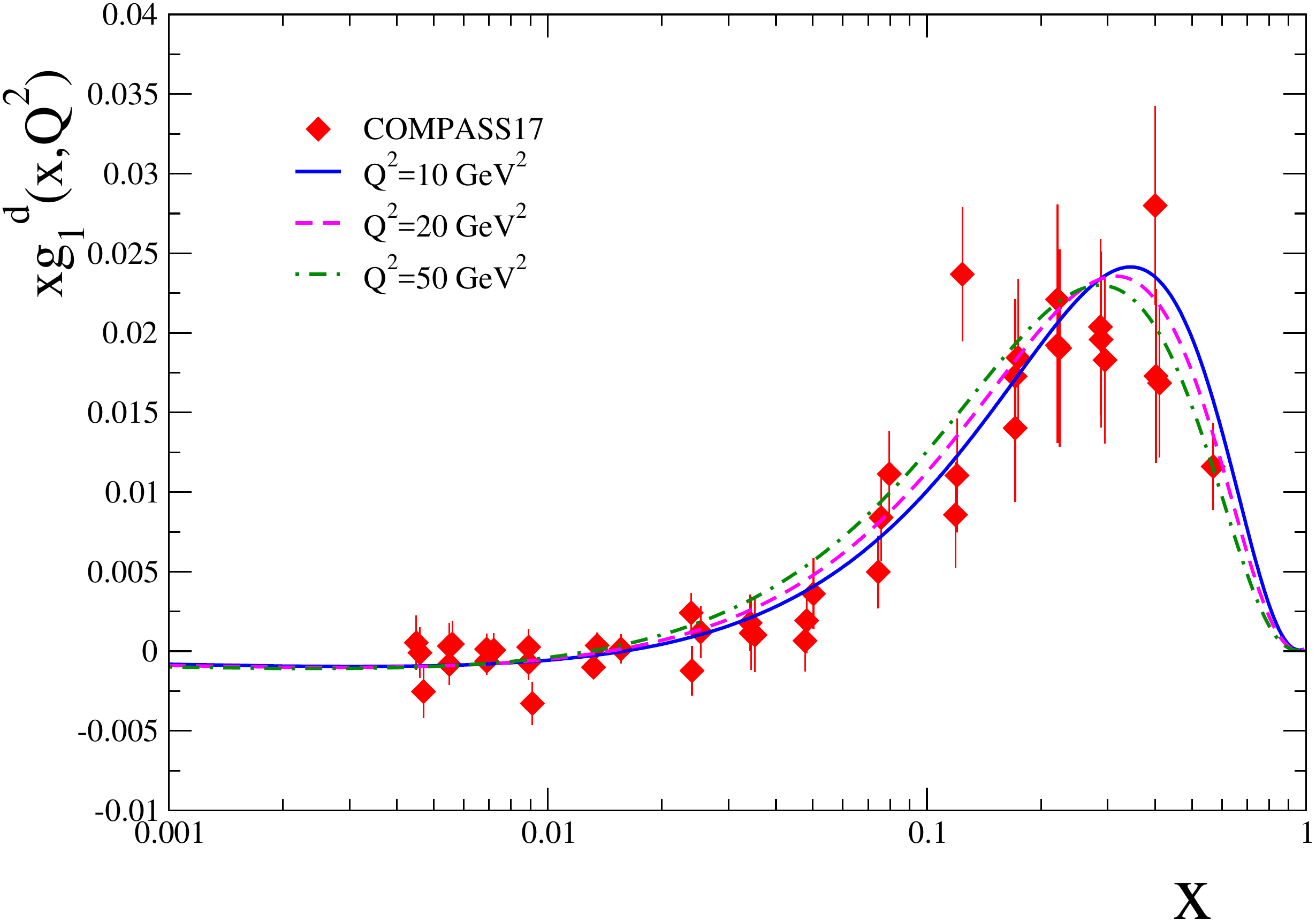}\caption{The \texttt{COMPASS17} \cite{Adolph:2016myg} data for the deuteron
structure function $xg_{1}^{d}(x,Q^{2})$ compared with our NLO results
(``Fit~B'') calculated at $Q^{2}$ = 10, 20, 50 GeV$^{2}$.{\small{}\label{fig:xg1d}}}
\end{figure}

To examine how the fits change with the inclusion of the COMPASS data,
we examine the partial $\chi^{2}$ contributions to COMPASS16{} and
COMPASS17{} data set for each of our fits: \{Base, Fit~A, Fit~B\}.
If we compute $\chi^{2}$ for the COMPASS16{} data set using the
``Base'' fit (which does not include this data), we find a total
$\chi^{2}$ value of 34.67 for the 51{} COMPASS16{} data points,
and when we include this data in the fit (``Fit~A'') it improves
slightly to 33.48. Correspondingly, if we fit the COMPASS17{} data
set using the ``Fit~A'' (which does not include this data), we find
a total $\chi^{2}$ value of 27.43 for the 43{} COMPASS17{} data
points, and in the fit (``Fit~B'') this is quite similar at 27.22.
Thus, both the COMPASS16{} and COMPASS17{} data set are in reasonable
agreement to the initial ``Base'' fit. The changes among the \{Base,
Fit~A, Fit~B\} sets is most evident in the ratio plots of Fig.~\ref{fig:ratio}.

Finally, in Figs.~\ref{fig:xg1p} and \ref{fig:xg1d}, we directly
compare our ``Fit~B'' with the proton and deuteron polarized structure
functions from \texttt{COMPASS16}~\cite{Adolph:2015saz} and \texttt{COMPASS17}~\cite{Adolph:2016myg}
experimental data in a composite plot; as the individual data range
over $Q^{2}$, we display our predictions with selected values of
$Q^{2}$ to illustrate the evolution effects. This allows us to see
the comparison of data and theory in a compact, albeit approximate,
manner. 

\subsubsection{$\alpha_{s}(Q^{2})$ Comparisons}

\begin{figure}[!t]
\begin{centering}
\includegraphics[clip,width=0.4\textwidth]{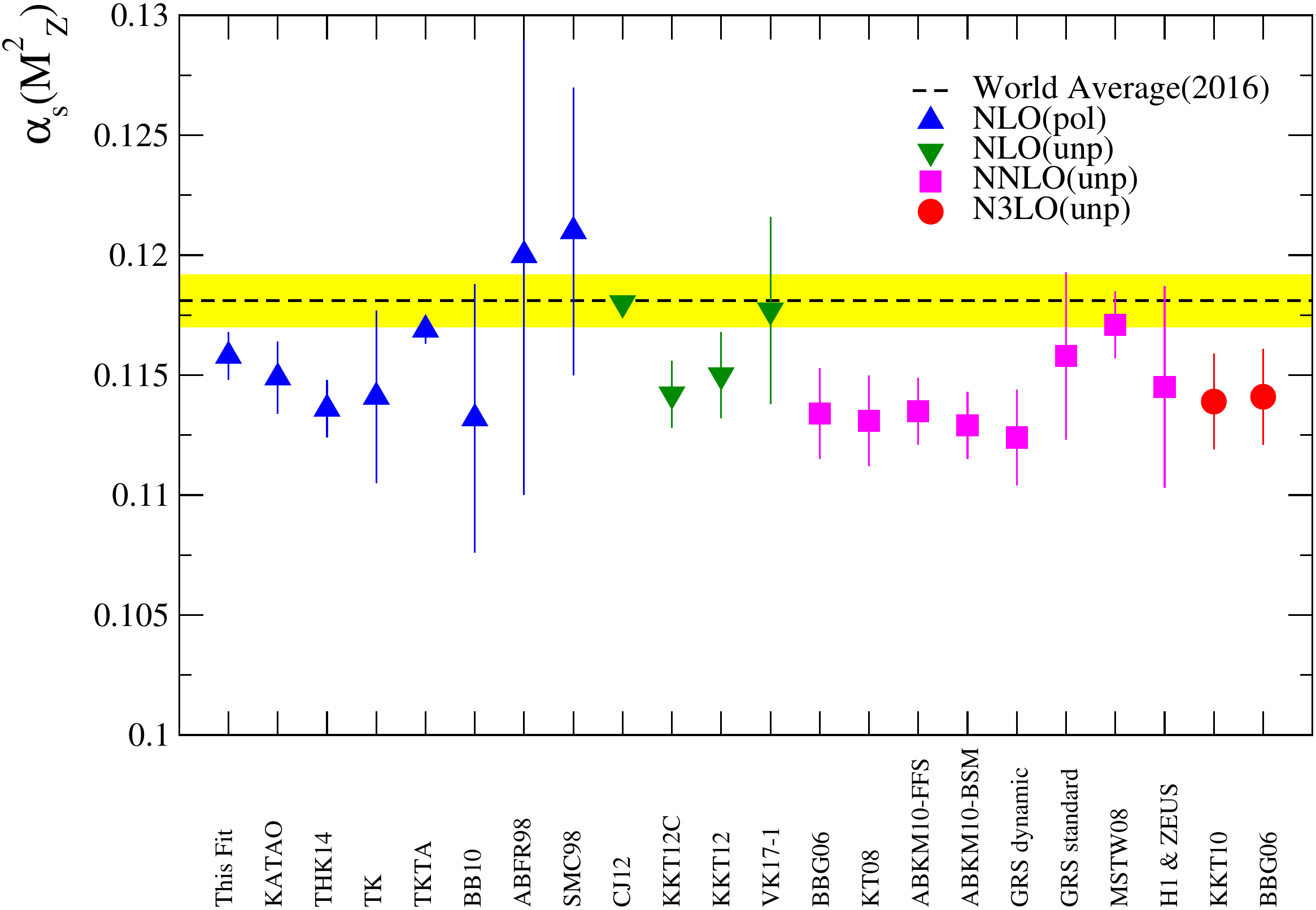} 
\par\end{centering}
\caption{The strong coupling constant $\alpha_{s}(M_{Z}^{2})$ values as compared
with different QCD analyses at NLO~\cite{Khorramian:2010qa,Shahri:2016uzl,Blumlein:2010rn,AtashbarTehrani:2007odq,Monfared:2014nta,Altarelli:1998nb,Ghosh:2000ys,Khanpour:2012tk,Adeva:1998vw,Vafaee:2017nze},
NNLO~\cite{Blumlein:2006be,Khorramian:2008yh,Alekhin:2009ni,Gluck:2006yz,Martin:2009bu,Aaron:2009aa,dEnterria:2015kmd},
and NNNLO~\cite{Blumlein:2006be,Khorramian:2009xz}. The dashed line
and yellow band shows the world average $\alpha_{s}(M_{Z}^{2})=0.1181\pm0.0011$
\cite{Patrignani:2016xqp}.{\small{}\label{fig:alphamz}}}
\end{figure}

In our present fits, we allowed $\alpha_{s}(Q_{0}^{2})$ to be a parameter
of the fit; these results are summarized in Table~\ref{tab:fit}.
We observe that the variation across our different fits is minimal, and
these values are consistent with the KATAO fit within uncertainties.
Although these values are extracted from data in the range $1\lesssim Q^{2}\lesssim100$,
we can evolve these up to $M_{Z}$ to compare with other values used
in the literature. Note that the $\alpha_{s}(Q^{2})$ evolution up to the
$M_{Z}^{2}$ scale will depend on the number of active flavors and
the mass scale of the transitions; we choose $m_{c}^{2}=3\,{\rm GeV}^{2}$
and $m_{b}^{2}=25\,{\rm GeV}^{2}$. Extrapolating our results up to
$M_{Z}$ at NLO order we find $\alpha_{s}(M_{Z}^{2})=0.1155$ for
Fit~A and $\alpha_{s}(M_{Z}^{2})=0.1158$ for Fit~B, and $\alpha_{s}(M_{Z}^{2})=0.1149$ for KATAO.
These values are low but within $2\sigma$ as compared to the world
average value of $\alpha_{s}(M_{Z}^{2})=0.1181\pm0.0011$~\cite{Patrignani:2016xqp},
and we display this in Fig.~\ref{fig:alphamz} along with various
results from the literature. 

\subsection{Moments and Sum Rules}

We now turn to integrated moments and sum rules. Note that the calculation
of the moments integrates over the full range $x=[0,1]$, so this
requires some extrapolation outside the $x$ range where the structure
functions have been measured.

\subsubsection{PPDF Moments}

\begin{table*}[!t]
\caption{{}Comparison of the first moments of the polarized parton
densities at NLO in the $\overline{{\rm MS}}$\textendash scheme at
$Q^{2}=4$ GeV$^{2}$. \label{tab:firstMom2}}
\begin{tabular}{lccccccc}
\hline 
 & ~~~~~~Base~~~~~~  & ~~~~~~Fit~A~~~~~~  & ~~~~~~Fit~B~~~~~~  & ~~~KATAO~\cite{Khorramian:2010qa}~~~  & ~~~BB \cite{Blumlein:2010rn}~~~  & ~~~GRSV~\cite{Gluck:2000dy}~~~  & ~~~AAC~\cite{Goto:1999by}~~~\\
\hline 
$\Delta u_{v}$  & $0.928$  & $0.928$  & $0.928$  & $0.928$  & $0.928$  & 0.9206  & 0.9278 \\
$\Delta d_{v}$  & $-0.342$  & $-0.342$  & $-0.342$  & $-0.342$  & $-0.342$  & \textendash 0.3409  & \textendash 0.3416 \\
$\Delta u$  & $0.873$  & $0.872$  & $0.876$  & $0.874$  & $0.866$  & 0.8593  & 0.8399 \\
$\Delta d$  & $-0.397$  & $-0.398$  & $-0.394$  & $-0.396$  & $-0.404$  & \textendash 0.4043  & \textendash 0.4295 \\
$\Delta\overline{q}$  & $-0.055$  & $-0.056$  & $-0.052$  & $-0.054$  & $-0.066$  & \textendash 0.0625  & \textendash 0.0879 \\
$\Delta g$  & $0.231$  & $0.161$  & $0.158$  & $0.224$  & $0.462$  & 0.6828  & 0.8076 \\
\hline 
\end{tabular}
\end{table*}

We start by computing the PPDF moments, as these will be the necessary
ingredients for the other moments and sum rules that follow.

In Table~\ref{tab:firstMom2}, we compare the results of the first
moments of the polarized parton densities for our fits with results
from the literature at NLO in the $\overline{{\rm MS}}$\textendash scheme
at $Q^{2}$ = 4 GeV$^{2}$. Comparing our ``Base'' fit with ``Fit
A'' and ``Fit~B'' we see the moments are generally stable with
the exception of the $\Delta g$ which varies by $\sim30\%$. Including
the other PPDF moments from the literature, we see the results for
$\{\Delta u_{v},\Delta d_{v}\}$ are quite stable ($\sim1\%$) while
$\{\Delta u,\Delta d\}$ show a bit more variation ($\sim10\%$),
and finally $\{\Delta\bar{q},\Delta g\}$ a larger spread ($>100\%$).
We will now look at the influence of the above PPDF moments on the
experimentally measurable structure functions.

\subsubsection{Structure Function Moments $\Gamma_{1}^{N}(Q^{2})$}

\begin{table}[t]
\caption{{}First moments of the polarized structure function \{$\Gamma_{1}^{p}$,
$\Gamma_{1}^{d}$, $\Gamma_{1}^{n}$, $\Gamma_{1}^{{\rm NS}}$\} for
``Fit~B'' at NLO at Q$^{2}$ = 3 GeV$^{2}$ compared with \texttt{{}COMPASS16}{}~\cite{Adolph:2015saz}
and\texttt{{} COMPASS17}{} \cite{Adolph:2016myg}
experimental data.~\label{table2:firstMomQ} }
\begin{tabular}{lccc}
\hline 
 & Fit~B  & \texttt{COMPASS16} {}\cite{Adolph:2015saz}  & \texttt{COMPASS17} {}\cite{Adolph:2016myg}\\
\hline 
$\Gamma_{1}^{p}$  & 0.133  & 0.139 $\pm$ 0.003 $\pm$ 0.009  & - \\
$\Gamma_{1}^{d}$  & 0.040  & -  & 0.043 $\pm$ 0.001 $\pm$ 0.003 \\
$\Gamma_{1}^{n}$  & -0.048  & -0.041 $\pm$ 0.006 $\pm$ 0.011  & - \\
$\Gamma_{1}^{{\rm NS}}$  & 0.182  & 0.181 $\pm$ 0.008 $\pm$ 0.014  & 0.192 $\pm$ 0.007 $\pm$ 0.015 \\
\hline 
\end{tabular}
\end{table}

\begin{table}[!t]
\caption{{}First moments of the polarized structure functions \{$\Gamma_{1}^{p}$,
$\Gamma_{1}^{d}$, $\Gamma_{1}^{n}$\} at $Q^{2}=5$ GeV$^{2}$ for
``Fit~B'' as compared to other results from the literature at NLO
in the $\overline{{\rm MS}}$\textendash scheme{}. \label{tab:firstMom4}}
\begin{tabular}{ccccc}
\hline 
 & Fit~B  & KATAO~\cite{Khorramian:2010qa}  & GRSV~\cite{Gluck:2000dy}  & AAC~\cite{Goto:1999by}\\
\hline 
$\Gamma_{1}^{p}$  & 0.135  & 0.133  & 0.132  & 0.137 \\
$\Gamma_{1}^{d}$  & 0.041  & 0.036  & 0.032  & 0.038 \\
$\Gamma_{1}^{n}$  & $-0.045$  & $-0.053$  & $-0.062$  & $-0.056$ \\
\hline 
\end{tabular}
\end{table}

We next examine the first moment of the $xg_{1}^{N}$ $(N=p,d,n)$
structure functions defined to be: 
\begin{equation}
\Gamma_{1}^{N}(Q^{2})\equiv\int_{0}^{1}g_{1}^{N}(x,Q^{2})dx\;.\label{eq:FirstMom}
\end{equation}
In Table~\ref{table2:firstMomQ}, we compare the results for $\Gamma_{1}^{N}(Q^{2})$
of Fit\,B with the COMPASS measurements. We observe that the fit agrees
with the COMPASS results within $\sim1\sigma$ of the experimental
uncertainty. 

Next, in Table~\ref{tab:firstMom4}, we compare our first moment
results with those from the literature. The theoretical results for
$\Gamma_{1}^{p}$ are uniform within $\pm2\%$, while the range on
$\Gamma_{1}^{d}$ increases to $\pm5\%$, and the range on $\Gamma_{1}^{n}$
further increases to $\pm15\%$.

\subsubsection{Bjorken Sum Rule, $xg_{1}^{NS}(x,Q^{2})$ and $\Gamma_{1}^{NS}(Q^{2})$ }

\begin{figure*}[!tbh]
\begin{centering}
\includegraphics[clip,width=0.4\textwidth]{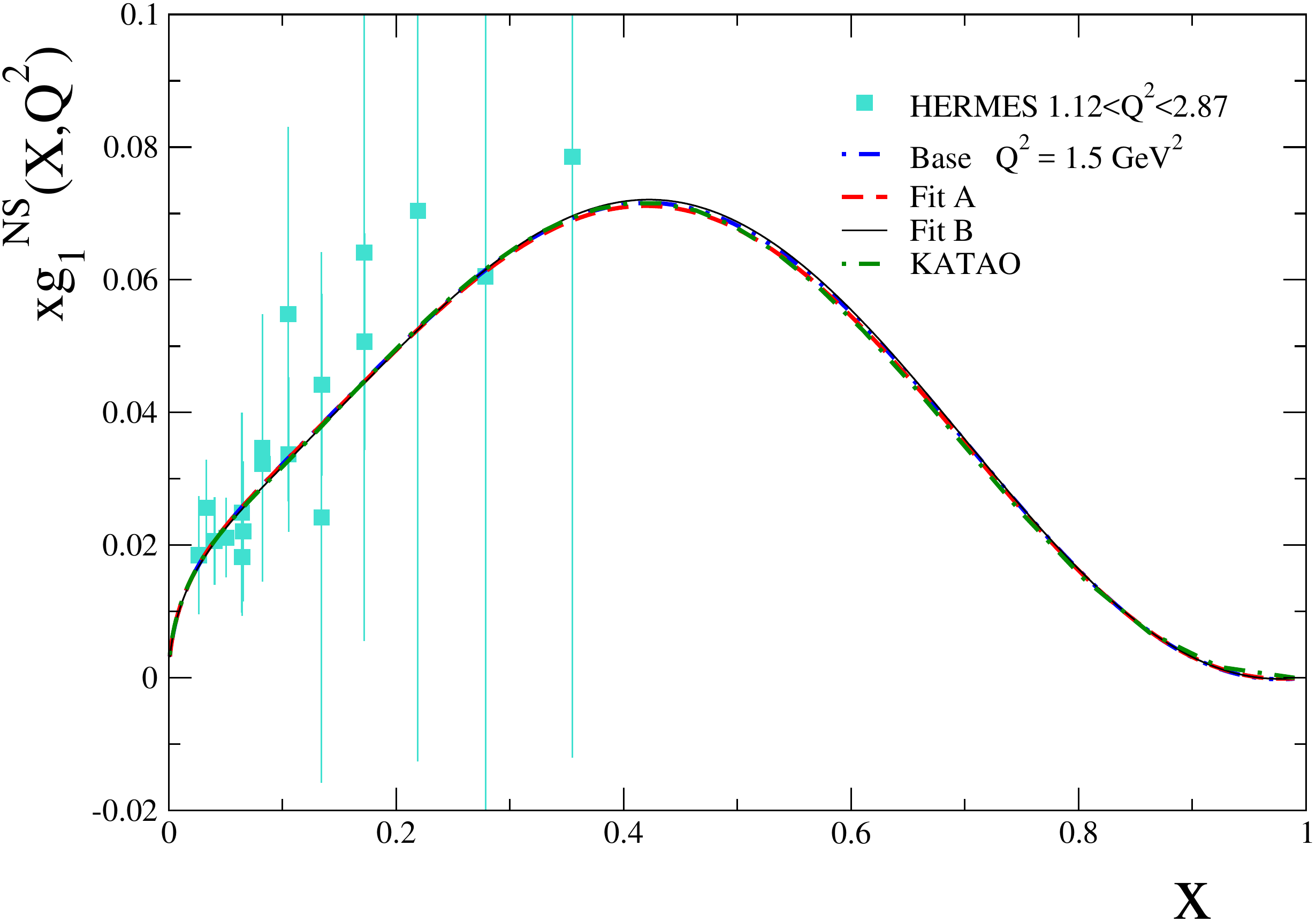}\hfill{}\includegraphics[clip,width=0.4\textwidth]{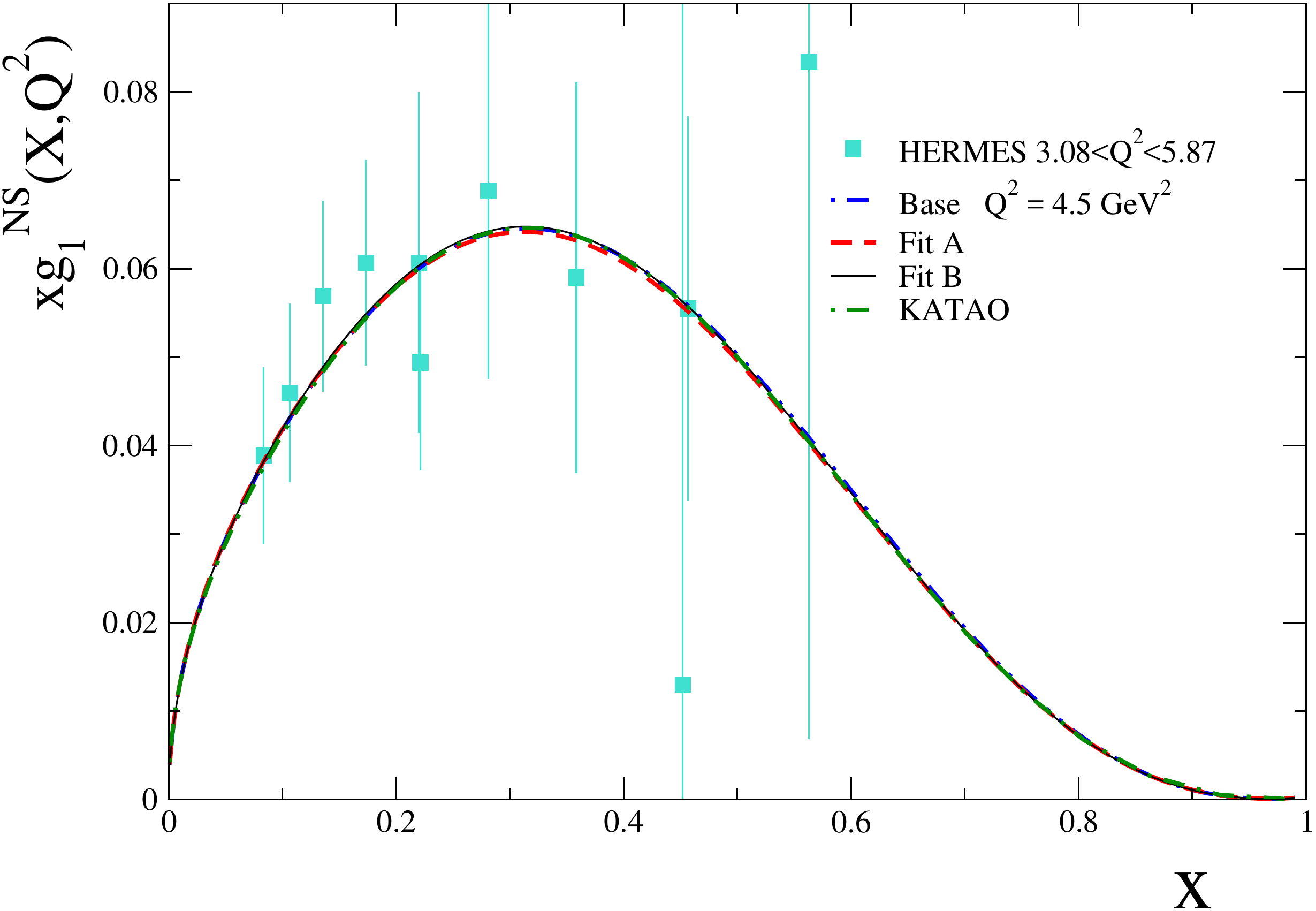} 
\par\end{centering}
\begin{centering}
\vspace{0.3cm}
 \includegraphics[clip,width=0.4\textwidth]{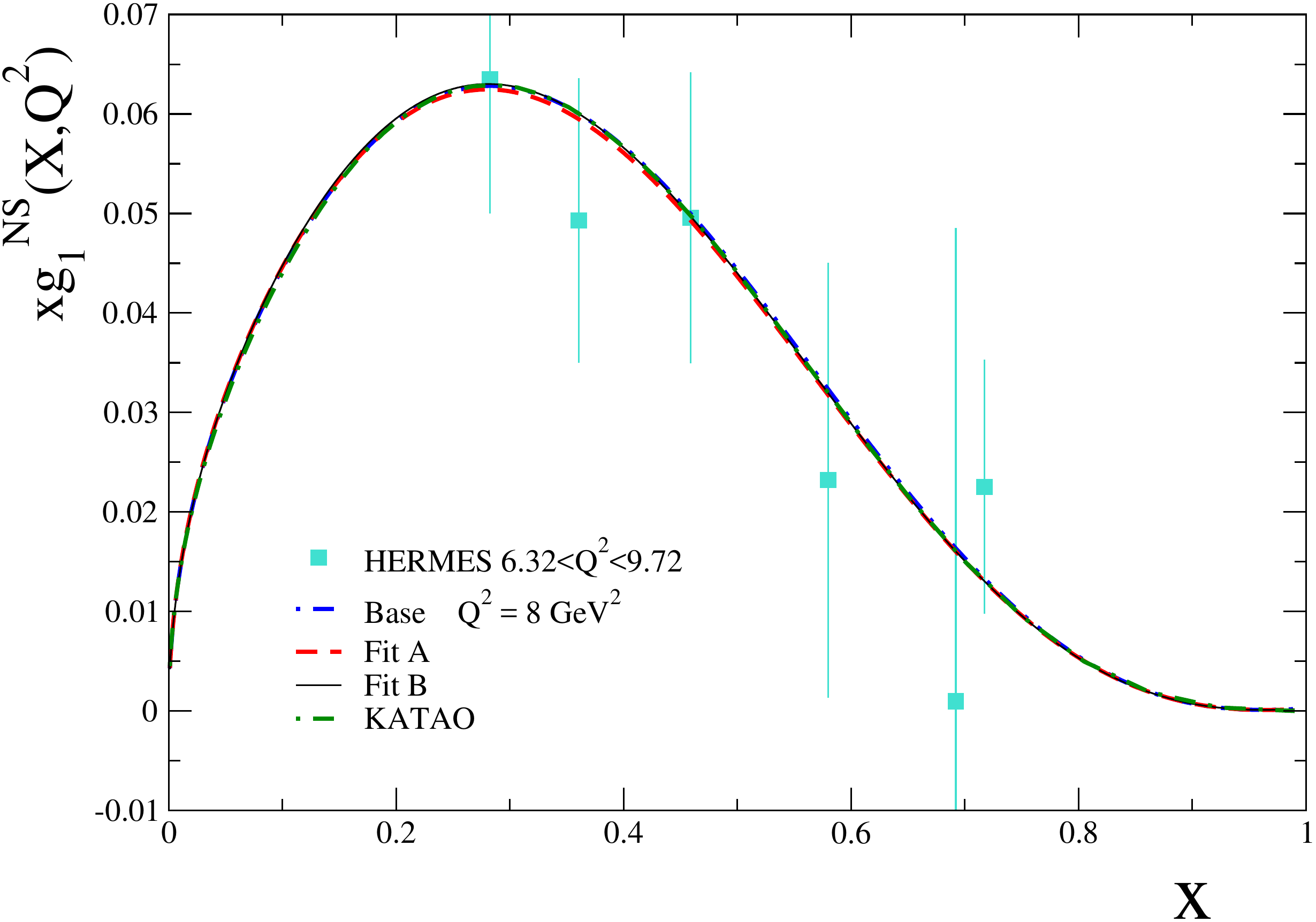}\hfill{}\includegraphics[clip,width=0.4\textwidth]{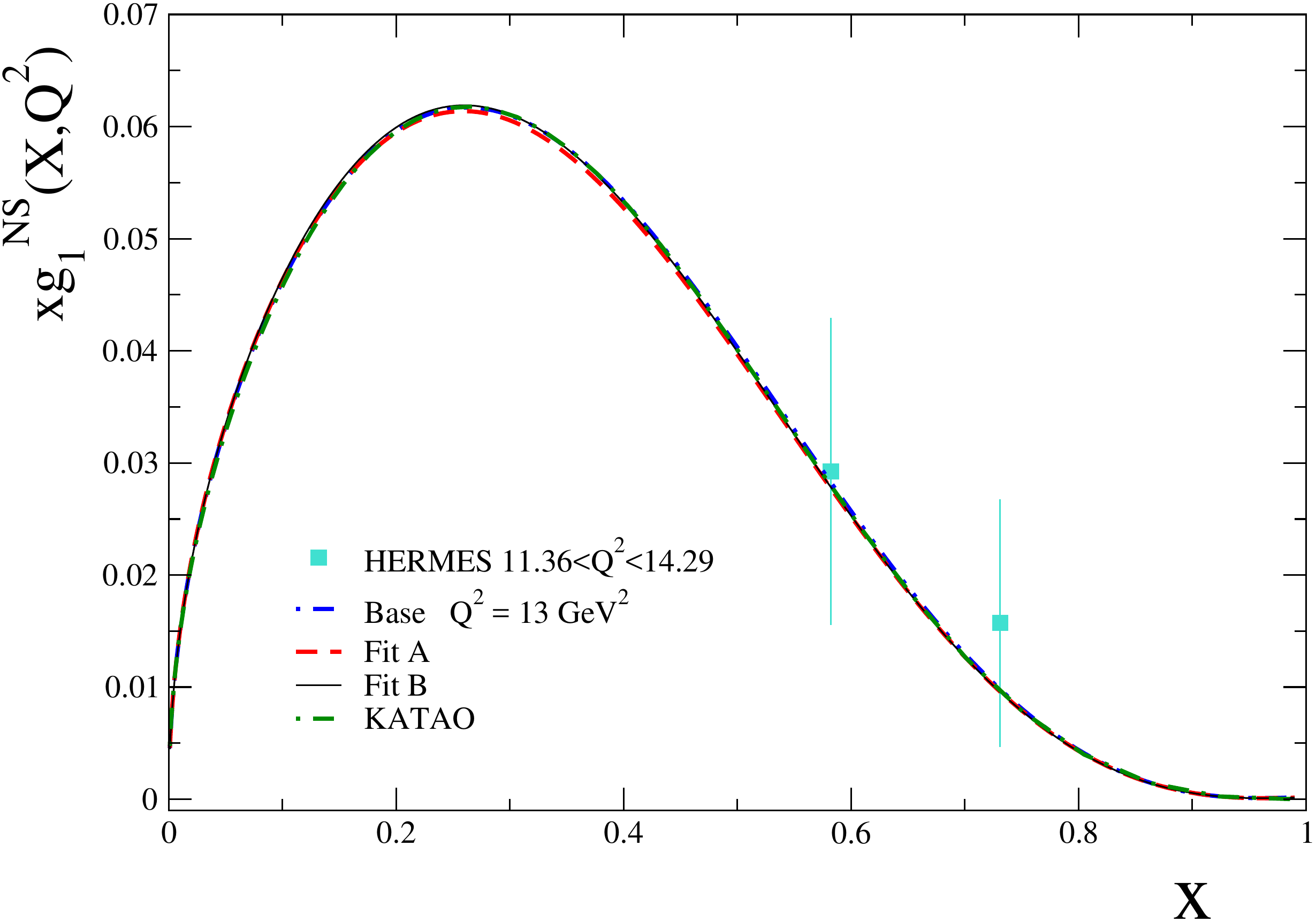} 
\par\end{centering}
\caption{NLO non-singlet polarized structure function $xg_{1}^{NS}(x,Q^{2})$
as function of $x$ in comparison with the results of KATAO~\cite{Khorramian:2010qa}
and HERMES experimental data \cite{Airapetian:2006vy}.{}\label{fig:g1NS}}
\end{figure*}

Following  Ref.~\cite{Airapetian:2006vy}, in the scaling (Bjorken) limit 
we have 
\begin{eqnarray}
\Gamma_1^{p,n}(Q^2) &=& \int_0^1  dx \, g_1^{p,n}(x,Q^2)  = \frac{1}{36}
\left( a_8 \pm 3 a_3 + 4 a_0 \right) \quad . 
\nonumber \\
\end{eqnarray}
We can isolate the $a_3$ term by taking the difference  
between the proton and neutron  terms, and we will
identify this as the non-singlet (NS) contribution. 
Thus, 
\begin{eqnarray}
\Gamma_{1}^{NS}(Q^{2}) &=& 
\Gamma_{1}^{p}(Q^{2}) -
\Gamma_{1}^{n}(Q^{2}) 
 =  \int_{0}^{1}g_{1}^{NS}(x,Q^{2})dx\nonumber \\
 & = & \frac{1}{6}|\frac{g_{A}}{g_{V}}|C_{1}^{NS}(Q^{2})~.
\end{eqnarray}
 $\Gamma_{1}^{NS}(Q^{2})$ enters the polarized Bjorken sum rule \cite{Bjorken:1969mm} 
 and is related to the ratio of the axial and
vector coupling constants ($g_{A,V}$). Here, $C_{1}^{NS}(Q^{2})$
is the non-singlet coefficient function.

In a similar manner we  define $g_{1}^{NS}(x,Q^{2})$ as
the difference between the proton and neutron structure functions: 

\begin{eqnarray}
xg_{1}^{NS}(x,Q^{2}) & \equiv & xg_{1}^{p}(x,Q^{2})-xg_{1}^{n}(x,Q^{2})\nonumber \\
 & = & 2[xg_{1}^{p}(x,Q^{2})-\frac{xg_{1}^{d}(x,Q^{2})}{1-\frac{3}{2}\omega_{D}}]~.\label{eq:g1ns}
\end{eqnarray}

In Fig.~\ref{fig:g1NS} we compare our results for $xg_{1}^{NS}(x,Q^{2})$
with the HERMES experimental data \cite{Airapetian:2006vy} for selected
bins of $Q^{2}$. We find minimal variation among our different theoretical
fits (including the previous KATAO fit), and these curves compare
well with the experimental results. 

From Eq.~(\ref{eq:g1ns}) we can also relate $\Gamma_{1}^{NS}(Q^{2})$
to the previously computed proton and neutron first moments as: 

\begin{eqnarray*}
\Gamma_{1}^{NS}(Q^{2}) & = & \Gamma_{1}^{p}(Q^{2})-\Gamma_{1}^{n}(Q^{2})\text{\quad}.
\end{eqnarray*}
These results are presented in Table~\ref{tab:firstMom4} and with
the \texttt{COMPASS} results. The result of our ``Fit~B'' is comparable
to \texttt{COMPASS16}, and below (but within uncertainties) to \texttt{COMPASS17}.

\subsubsection{$g_{2}^{N}(x,Q^{2})$ Structure Functions}

We can also calculate the structure function $g_{2}^{N}(x,Q^{2})$
via the Wandzura-Wilczek relation \cite{Wandzura:1977qf,Piccione:1997zh}:

\begin{equation}
g_{2}^{N}(x,Q^{2})=-g_{1}^{N}(x,Q^{2})+\int_{x}^{1}\frac{dy}{y}g_{1}^{N}(y,Q^{2})~.\label{eq:xg2}
\end{equation}

\begin{figure}[!t]
\begin{centering}
\includegraphics[clip,width=0.4\textwidth]{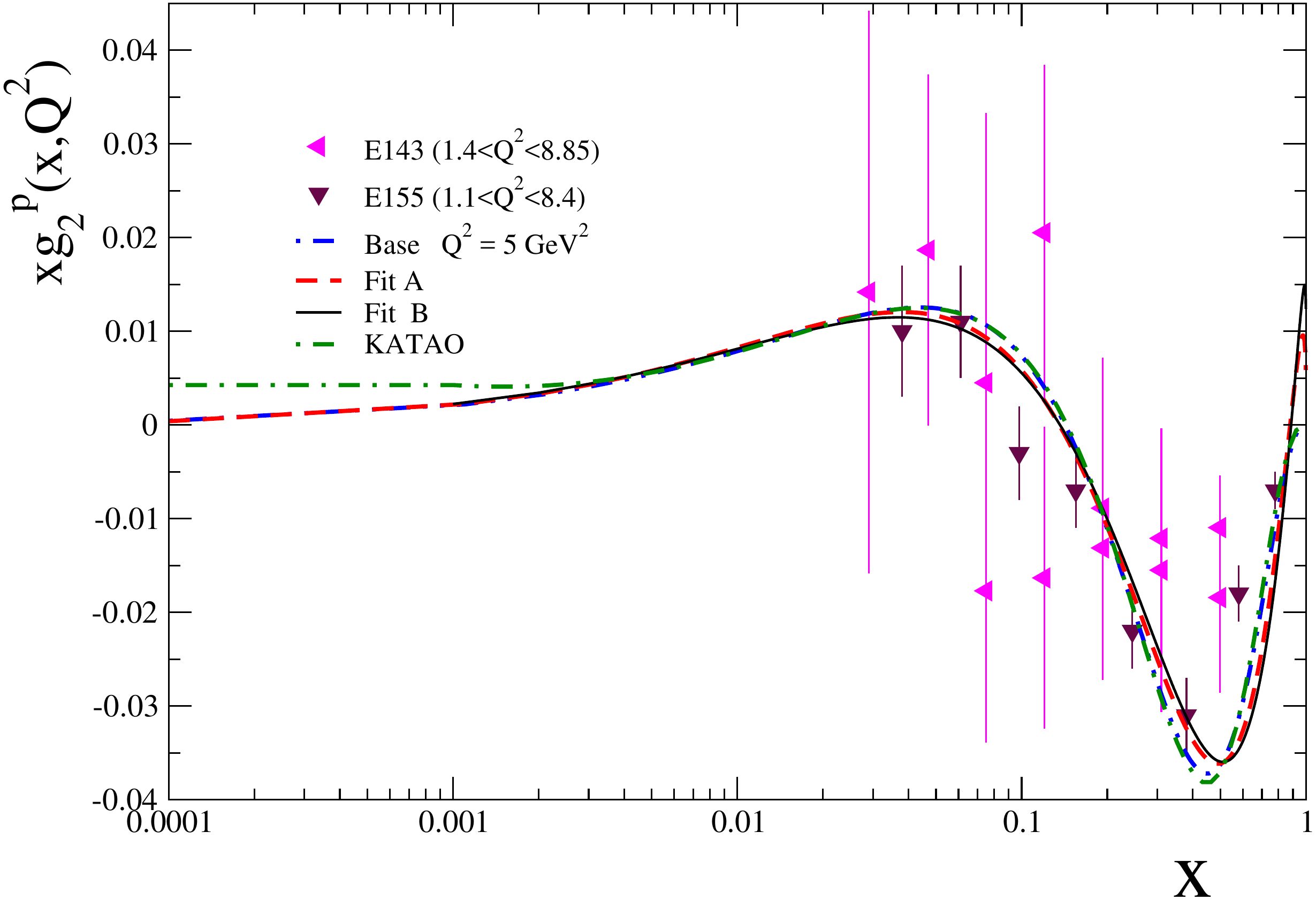} \includegraphics[clip,width=0.4\textwidth]{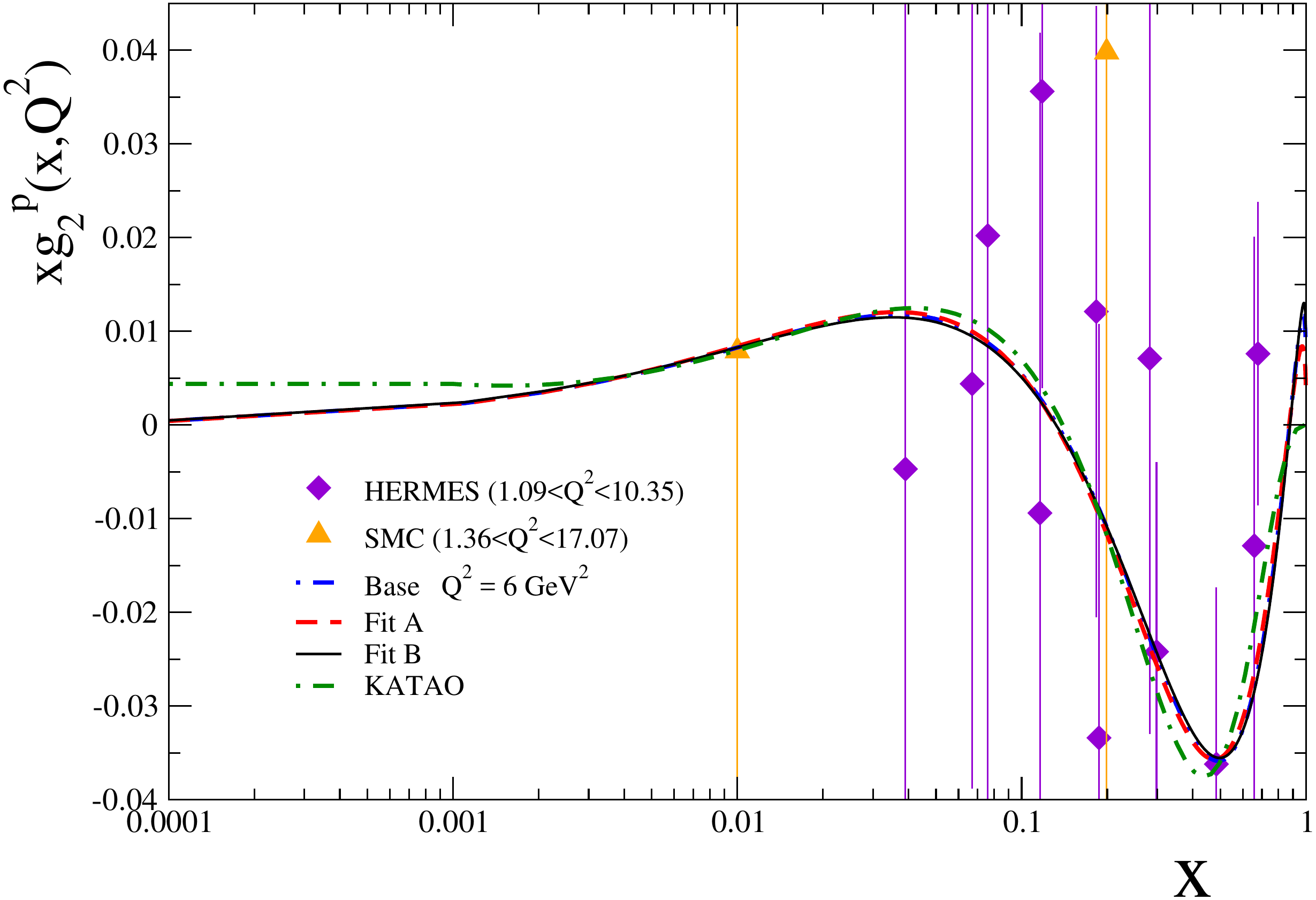}
\includegraphics[clip,width=0.4\textwidth]{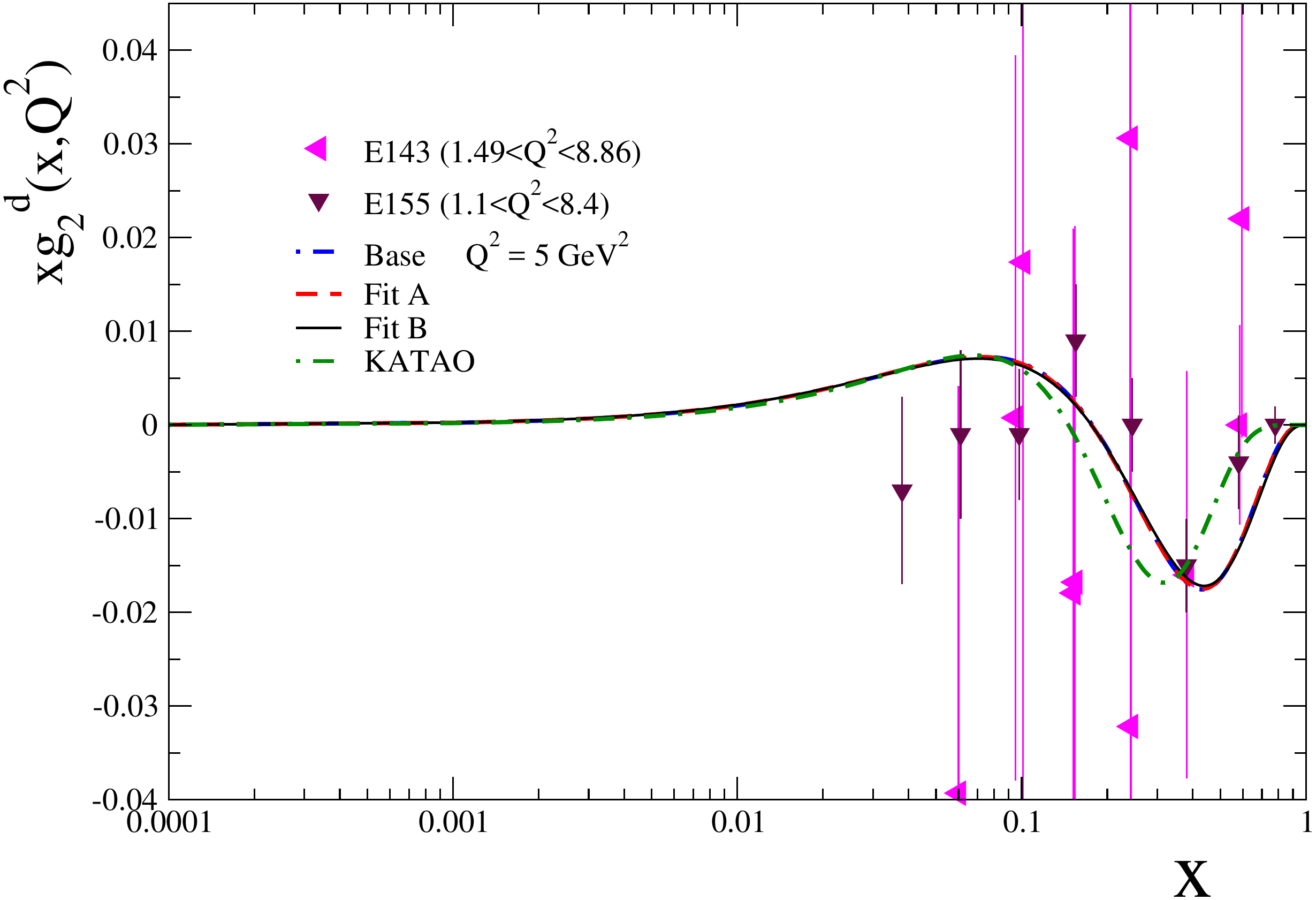} 
\par\end{centering}
\caption{NLO polarized structure function $xg_{2}^{p}(x,Q^{2})$ and $xg_{2}^{d}(x,Q^{2})$
as a function of $x$ for $Q{^{2}}$ = 5, 6 GeV$^{2}$ compared to
E143 \cite{Abe:1998wq}, E155 \cite{Anthony:2002hy}, HERMES \cite{Airapetian:2011wu}, and
SMC \cite{Adams:1997tq} experimental data. We present also the results
of different Base (dashed-dotted-dotted), Fit~A (dashed) and Fit~B
(solid line) QCD fits which are compared with our previous KATAO (dashed-dotted)
results. \label{fig:g2pg2d}}
\end{figure}

\begin{figure}[!t]
\begin{centering}
\includegraphics[clip,width=0.4\textwidth]{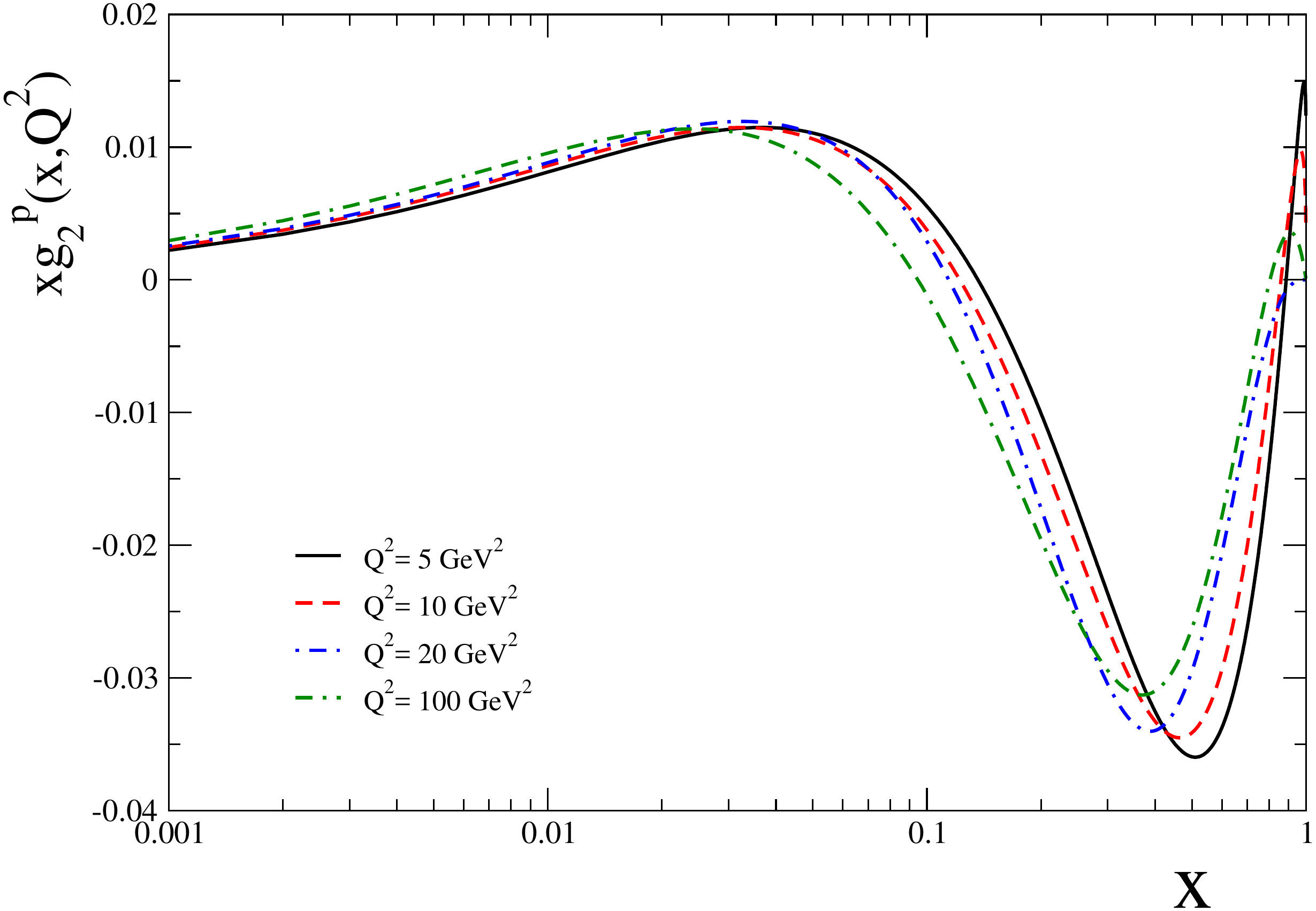} \includegraphics[clip,width=0.4\textwidth]{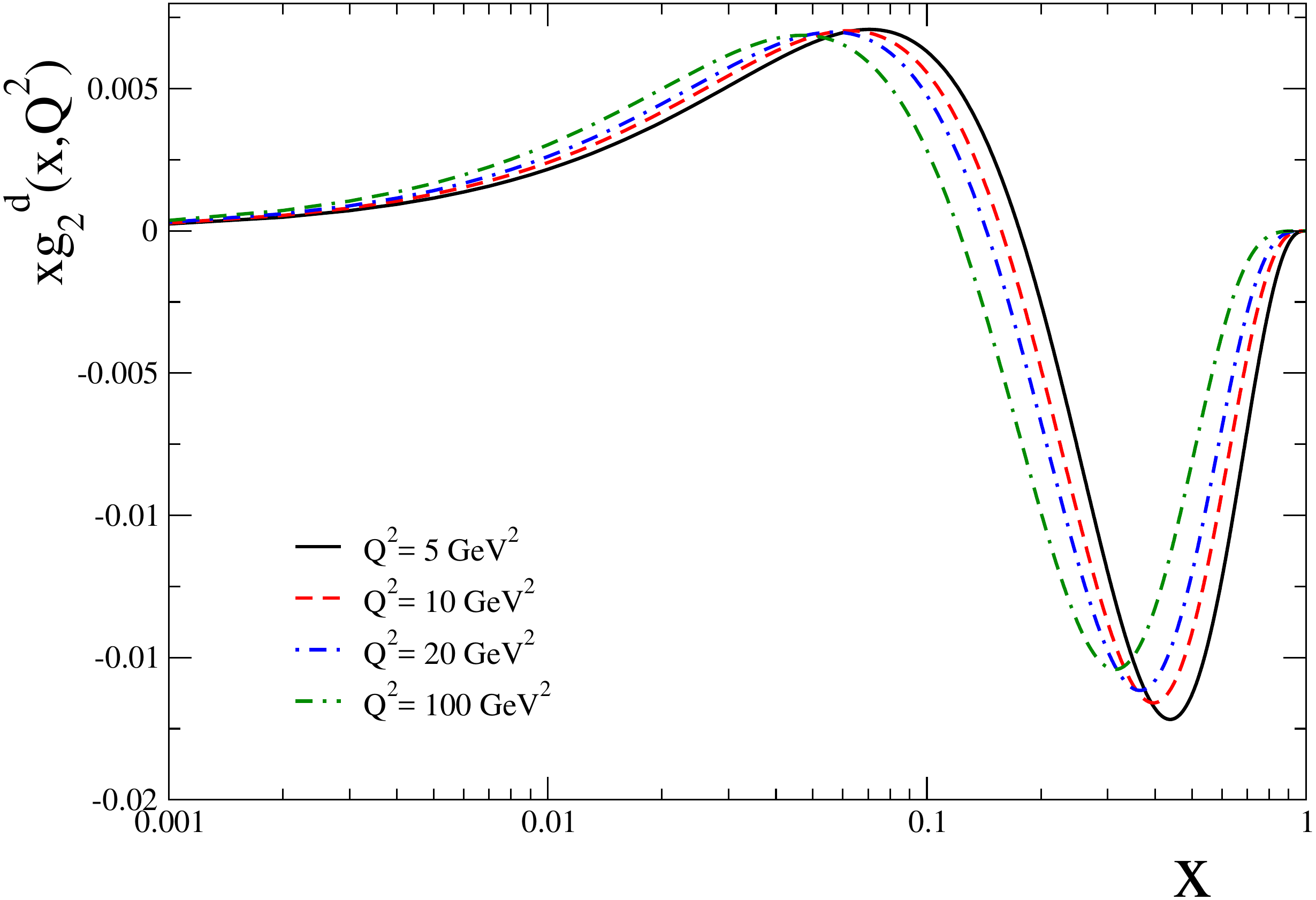} 
\par\end{centering}
\caption{NLO polarized structure function $xg_{2}(x,Q^{2})$ for the proton
and deuteron as function of $x$ and for Q$^{2}$ = 5, 10, 20, 100
GeV$^{2}$.{\small{}\label{fig:g2pd}}}
\end{figure}
Figure \ref{fig:g2pg2d} shows the polarized structure function $xg_{2}^{p}$
and $xg_{2}^{d}$ as a function of $x$ for different cases of Base,
Fit~A, Fit~B and our previous KATAO results \cite{Khorramian:2010qa}
in comparison with E143 \cite{Abe:1998wq}, E155 \cite{Anthony:2002hy},
HERMES \cite{Airapetian:2011wu}, and SMC \cite{Adams:1997tq} experimental data at $Q^{2}$ = 5
, 6 GeV$^{2}$. As the data actually span over a range of $Q^{2}$,
in Fig. \ref{fig:g2pd} we display the $Q^{2}$ evolution of the
polarized structure function $xg_{2}(x,Q^{2})$ for the proton and
deuteron as function of $x$. In Fig. \ref{fig:g2pg2d} we see that
our ``Base'' and ``Fit~A'' coincide throughout the $x$ range
suggesting a minimal impact from the\texttt{ COMPASS16} data on this
observable; conversely, our ``Fit~B'' does differ, especially in
the larger $x$ region, suggesting a stronger influence of the\texttt{
COMPASS17} data on $xg_{2}^{d}(x,Q^{2})$.

\subsection{The Proton Spin }

\begin{table}[!t]
\caption{{}Spin contribution of the proton in the NLO approximations at
$Q^{2}=4$ GeV$^{2}$ for Base, Fit~A, Fit~B compared with the KATAO~\cite{Khorramian:2010qa}.
{}We have computed $\{\frac{1}{2}\Delta\Sigma,\Delta g\}$ using
our PPDFs, and inferred $L_{z}$ assuming precisely $\nicefrac{1}{2}$
for the proton spin.{} \label{tab:sumrule}}
{} 

\begin{tabular}{lcccc}
\hline 
 & ~~~$\frac{1}{2}\Delta\Sigma$~~~  & ~~~$\Delta g$~~~  & ~~~$L_{z}$~~~  & $\frac{1}{2}\Delta\Sigma+\Delta g+L_{z}$\\
\hline 
KATAO  & $0.131$  & $0.224$  & $0.145$  & $\nicefrac{1}{2}$ \\
\hline 
Base  & $0.129$  & $0.231$  & $0.140$  & $\nicefrac{1}{2}$\\
Fit~A  & $0.125$  & $0.161$  & $0.214$  & $\nicefrac{1}{2}$\\
Fit~B  & $0.139$  & $0.158$  & $0.203$  & $\nicefrac{1}{2}$\\
\hline 
\end{tabular}
\end{table}
\begin{figure}[!t]
\begin{centering}
\includegraphics[clip,width=0.4\textwidth]{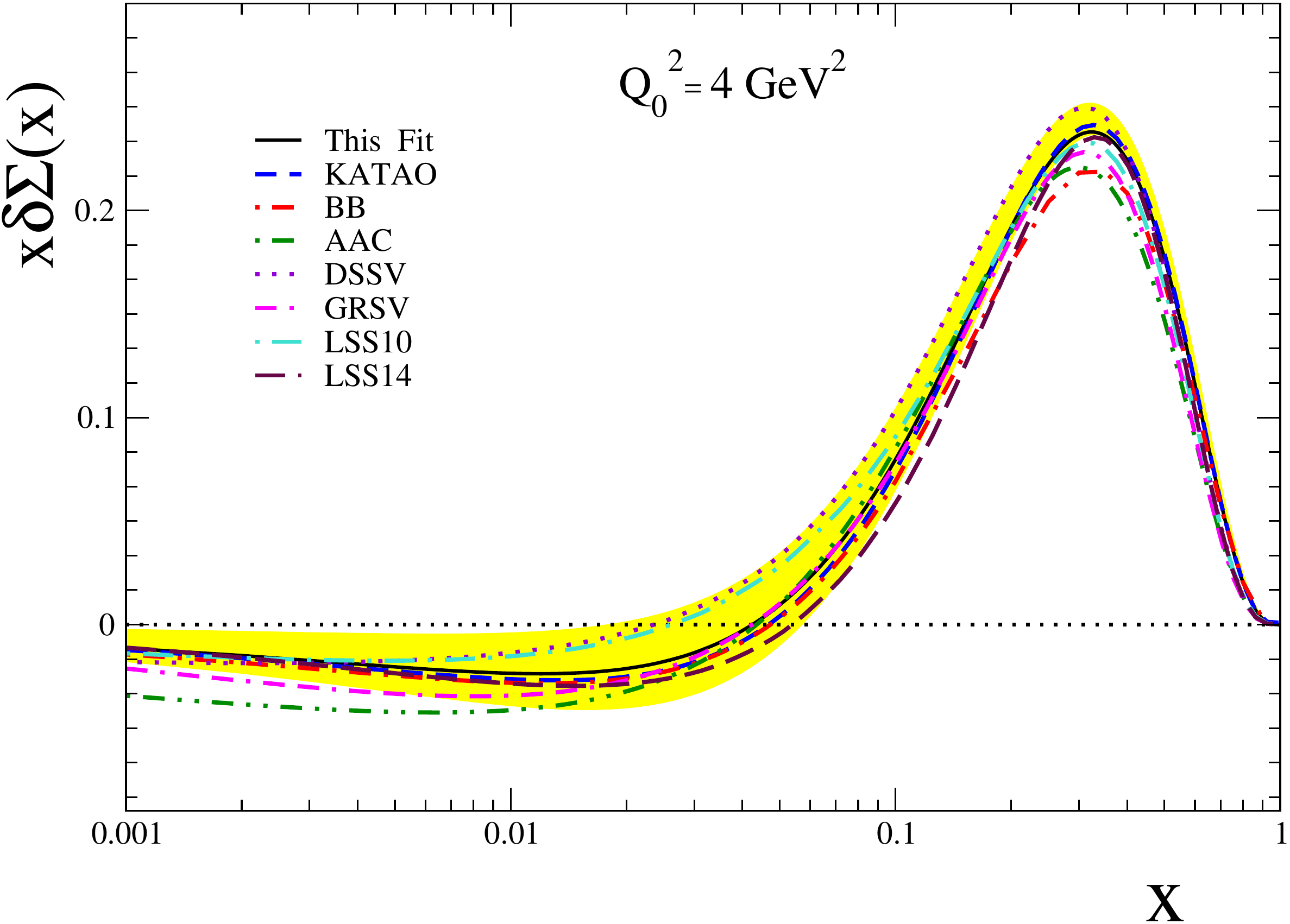} 
\par\end{centering}
\caption{NLO polarized singlet parton density $x\delta\Sigma(x)$ at $Q_{0}^{2}$
= 4 GeV$^{2}$ for this fit (``Fig B'') as a function of $x$, compared
with results from the literature including KATAO~\cite{Khorramian:2010qa},
BB~\cite{Blumlein:2010rn}, AAC~\cite{Goto:1999by}, DSSV~\cite{deFlorian:2008mr},
GRSV~\cite{Gluck:2000dy} and LSS~\cite{Leader:2010rb,Leader:2014uua}{}. \label{fig:sigma}}
\end{figure}

It is important for us to understand the decomposition of the proton
spin in terms of the separate contributions from the quarks, gluon,
and the orbital angular momentum components. The spin of the proton
can be computed from the first moment of the polarized parton densities
together with the quark and gluon orbital momentum ($L_{q},L_{g}$)
is as following \cite{Leader:2016sli} 
\begin{eqnarray}
\frac{1}{2}=\frac{1}{2}\Delta\Sigma(Q^{2})+\Delta g(Q^{2})+{\mathrm{L}}_{z}(Q^{2})~.\label{spinSR}
\end{eqnarray}
Here ${\mathrm{L}}_{z}(Q^{2})={\mathrm{L}}_{q}(Q^{2})+{\mathrm{L}}_{g}(Q^{2})$
is the total orbital angular momentum of all the quarks and gluons,
$\Delta{g(Q^{2})}=\int_{0}^{1}dx~\delta g(x,Q^{2})$ is the first
moment of the polarized gluon distribution, and $\Delta\Sigma(Q^{2})=\int_{0}^{1}dx~\delta\Sigma(x,Q^{2})$
with $\delta\Sigma\equiv\delta u_{v}+\delta d_{v}+6\delta{\bar{q}}$
is the first moment of the polarized singlet distribution. In Eq.~(\ref{spinSR}),
we note that the spin sum ($\nicefrac{1}{2}$) is actually independent
of $Q^{2}$ even though each individual term is dependent on $Q^{2}$.

In Table~\ref{tab:sumrule} we compute $\{\nicefrac{1}{2}\Delta\Sigma,\Delta g\}$
using ``Fit~B'' at $Q^{2}=4$ GeV$^{2}$, and then infer the value
of ${\mathrm{L}}_{z}(Q^{2})$ assuming Eq.~(\ref{spinSR}). As we
observed in Table~\ref{tab:firstMom2} the values for $\nicefrac{1}{2}\Delta\Sigma$
show minimal variation while there is a larger spread for $\Delta g(Q^{2})$
which then implies a larger spread of $L_{z}$.

The comparison of ``Fit~B'' with other $x\delta g(Q^{2})$ from
the literature were displayed in Fig.~\ref{fig:pdf11}, and there
is quite a bit of variation. In contrast, Figure \ref{fig:sigma}
shows our NLO singlet polarized parton density $x\delta\Sigma(x)$
($\equiv\delta u_{v}+\delta d_{v}+6\delta{\bar{q}}$) compared with
other results from the literature. The results of this fit (``Fit
B'') with the previous analysis KATAO \cite{Khorramian:2010qa} are
quite similar as suggested by Table~\ref{tab:firstMom2}. Generally,
the singlet polarized distributions are negative for $x\lesssim0.04$
for most of the models, but there is some slight variation in the
range $x\lesssim0.06$ to $x\lesssim0.02$. Overall, the variation
of $x\delta\Sigma(x)$, as compared to $x\delta g(Q^{2})$, is reduced;
this is notable as $x\delta\Sigma(x)$ is a combination of both valence
and sea PPDFs.

\section{Conclusions }

We performed a QCD analysis of the deep inelastic nucleon scattering
data from COMPASS \cite{Alekseev:2010hc,Ageev:2005gh,Alexakhin:2006oza},
HERMES \cite{Ackerstaff:1997ws,Airapetian:1998wi,Airapetian:2006vy},
SLAC \cite{Abe:1998wq,Anthony:2000fn,Anthony:1999rm,Ashman:1989ig,Anthony:1996mw,Abe:1997cx},
EMC \cite{Ashman:1987hv}, and SMC \cite{Adeva:1998vv} at NLO. This
also included the recent data from \texttt{COMPASS16} \cite{Adolph:2015saz}
and \texttt{COMPASS17} \cite{Adolph:2016myg} for the proton and deuteron
polarized structure function measurements. 

We extracted the PPDFs and $\alpha_{s}(Q_{0}^{2})$ with uncertainties
using a $\chi^{2}$ minimization, and compared our results with those
from the literature including AAC~\cite{Goto:1999by}, DSSV \cite{deFlorian:2008mr}, BB~\cite{Blumlein:2010rn},
GRSV~\cite{Gluck:2000dy}, LSS~\cite{Leader:2010rb,Leader:2014uua}, and KATAO~\cite{Khorramian:2010qa}.
In contrast to our previous polarized analysis (KATAO), we did not
use the Jacobi polynomial expansion method. Our results for the PPDFs
are comparable to other extractions, and generally it appears that
$x\delta u_{v}$ and $x\delta d_{v}$ are comparatively well determined
in contrast to $x\delta\bar{q}$ and $x\delta g$ which display a
larger variation across the $x$ range.

We also computed various structure functions and moments for the proton,
neutron, and deuteron, and these also compare well with the both the
\texttt{COMPASS} data, as well as other determinations from the literature.
Again the results from this fit are comparable to the previous KATAO~\cite{Khorramian:2010qa}
results using orthogonal polynomials; it is reassuring to see that the
results are generally independent of the underlying calculational
methodology. 

The strong coupling constant $\alpha_{s}(Q_{0}^{2})$ was extracted
from the fits, and the uncertainty is slightly decreased compared
to the KATAO analysis. This $\alpha_{s}(Q_{0}^{2})$ can be evolved
up to $\alpha_{s}(M_{Z}^{2})$ by assuming an evolution order (LO,
NLO, ...) and heavy quark mass thresholds; we find values that are
low compared to the world average, but within uncertainties. 

From this analysis, it appears that both the various theoretical analyses
using a variety of techniques and ($x$-space, $N$-space, orthogonal
polynomials) are generally converging to yield a homogeneous set of
predictions which are in good agreement with the diverse sets of experimental
measurements. While there is still room for further improvements,
such studies provide a strong validation of the underlying QCD theoretical
framework.

A standard \texttt{LHAPDF} library file of our polarized PDFs \{$x\delta u_{v}(x,Q^{2})$,
$x\delta d_{v}(x,Q^{2})$, $x\delta\bar{q}(x,Q^{2})$, $x\delta g(x,Q^{2})$\}
and their uncertainties can be obtained via e-mail from the authors upon request. 

\section*{Acknowledgments}

We gratefully acknowledge O. Denisov and E. Kabuss of the COMPASS
collaboration, as well as D. Stamenov and S. Bass for detailed comments and helpful discussions and suggestions.  We also thank F. Arbabifar and M. Soleymanini for useful comments. A. K. is grateful to the CERN TH-PH division for their hospitality where a portion of this work was performed. The work of F.O.{} was supported in part
by the U.S.{} Department of Energy under Grant No. DE-SC0010129.


 \bibliographystyle{apsrev}
\bibliography{paper}

\end{document}